\documentclass[preprint,showpacs,tightenlines,amsmath,amssymb]{revtex4}
\usepackage{graphicx}
\def \beq {\begin{equation}}
\def \eeq {\end{equation}}
\def \beqa {\begin{eqnarray}}
\def \eeqa {\end{eqnarray}}
   
\newcommand{\vect}[1]              
           {\mbox{\boldmath$#1$}}  
\begin{document}
%
\title{Sum Rule Approach to the Isoscalar Giant Monopole\\
Resonance in Drip Line Nuclei}
\author{M. Centelles$^1$, X. Vi\~nas$^1$, S. K. Patra$^2$, 
J. N. De$^{3,4}$ and Tapas Sil$^1$}
\affiliation{$^1$Departament d'Estructura i Constituents de la Mat\`eria,\\
     Facultat de F\'{\i}sica, Universitat de Barcelona,
     Diagonal {\sl 647}, E-{\sl 08028} Barcelona, Spain\\
$^2$Institute of Physics, Sachivalaya Marg,
     Bhubaneswar-{\sl 751 005}, India\\
$^3$Variable Energy Cyclotron Centre, 
1/AF, Bidhannagar, Kolkata-700064, India\\
$^4$Saha Institute of Nuclear Physics, 
1/AF, Bidhannagar, Kolkata-700064, India}
%
\begin{abstract}
Using the density-dependent Hartree-Fock approximation and Skyrme
forces together with the scaling method and constrained Hartree-Fock
calculations, we obtain the average energies of the isoscalar giant
monopole resonance. The calculations are done along several isotopic
chains from the proton to the neutron drip lines. It is found that
while approaching the neutron drip line, the scaled and the
constrained energies decrease and the resonance width increases.
Similar but smaller effects arise near the proton drip line, although
only for the lighter isotopic chains. A qualitatively good agreement
is found between our sum rule description and the presently existing
random phase approximation results. The ability of the semiclassical
approximations of the Thomas-Fermi type, which properly describe the
average energy of the isoscalar giant monopole resonance for stable
nuclei, to predict average properties for nuclei near the drip lines
is also analyzed. We show that when $\hbar$-corrections are included,
the semiclassical estimates reproduce, on average, the quantal
excitation energies of the giant monopole resonance for nuclei with
extreme isospin values.
\end{abstract}
\maketitle
%
\section{Introduction}

Experimental and theoretical studies of exotic nuclei with extreme
isospin values are presently one of the more active areas of research in
nuclear physics. Recent developments in accelerator technology and
detection techniques allow research beyond the limits of
$\beta$-stability. The number of unstable nuclei for which masses have
been measured is rapidly increasing \cite{R1} and this trend is
expected to be continued due to the use of radioactive beams
\cite{R1a,R1b,R1c}. In particular, the proton drip line has been
approached as far as for Pb isotopes \cite{R1d}. The neutron drip line
has been reached to date for $Z \le 8$ and is expected to be extended
to $Z \le 28$ in the present decade \cite{R1e}.

Collective phenomena are very useful tools to study the nuclear
structure as well as to test the ability of the nuclear effective
forces in describing such situations. The analysis of the small
amplitude oscillations, i.e. the giant resonances, is of special
relevance. In particular, it is very interesting to study the
isoscalar giant monopole resonance (ISGMR) from where the
incompressibility modulus of nuclear matter ($K_{\infty}$) can be
extracted \cite{R14,R2}. The value of $K_{\infty}$ is an important
ingredient not only for the description of finite nuclei but also for
the study of heavy ion collisions, supernovae, and neutron stars.

The ISGMR for stable nuclei has been studied long ago from both,
experimental and theoretical standpoints \cite{R4}. The basic theory
for the microscopic description of these collective motions is the
Random Phase Approximation (RPA) \cite{R5,R6}. The RPA calculations
allow to obtain the strength distribution $S(E)$ which measures the
response of a nucleus to an external perturbation. In the case of the
giant resonances, $S(E)$ is usually concentrated in a rather narrow
region of the energy spectrum, at least for heavy stable nuclei. Thus
the knowledge of a few low energy-weighted moments of $S(E)$ (sum
rules) can provide a useful information on the average properties of
the giant resonances, as for instance the energy of the centroids and the
resonance widths. The full RPA calculation can be avoided by using the
so-called sum rule approach in which several selected odd moments of
$S(E)$ are obtained by means of the properties of the ground state
only \cite{R11} and thus used to evaluate these average properties.
However, the full quantal calculation of the sum rules is still a
complicated task. In some particular cases, it can be simplified by
using the scaling method \cite{R11,R10,R12} to obtain the cubic energy
weighted moment or performing constrained Hartree-Fock (HF)
calculations \cite{R12,R11,R13,R14} which allow to compute the inverse
energy weighted moment.

Nuclei near the drip lines are characterized by the small energy of
the last bound nucleons and by their large asymmetry $I=(N-Z)/A$,
therefore it is expected that their properties, in particular the ones
of the collective excitations, may considerably differ from the
corresponding properties of the stable nuclei. The theoretical
analysis of the giant monopole resonances of some exotic nuclei has
been worked out in the last few years. For instance the RPA formalism
together with Skyrme forces have been used to study the ISGMR of some
isotopes of $A=$100, 110 and 120 \cite{R28} and of some Ca isotopes
\cite{R15,R15a}. These RPA calculations give us a detailed information
about the specific nuclei considered. The qualitative behavior of the
ISGMR of other exotic nuclei may be inferred from these RPA results;
however, a wider study of the properties of the collective modes near
the drip lines is still lacking. In order to obtain a global insight
into the behavior of the ISGMR near the drip lines, we will study in
this paper the aforementioned average properties of this collective
excitation using the sum rule approach along different isotopic chains
covering the whole periodic table. The scaling transformation of the
density applied together with zero-range Skyrme forces allows to
express, for the monopole oscillations, the $m_3$ RPA sum rule (i.e.
the cubic weighted moment of the strength distribution) by a simple
and closed formula related to different contributions to the
ground-state energy \cite{R11}. Spherical constrained HF calculations
also allow to obtain the inverse energy weighted moment of the
strength distribution (the RPA $m_{-1}$ sum rule) for the monopole
oscillation in an easy way \cite{R11}. Although with the sum rule
approach a detailed information on the RPA strength cannot be
obtained, this method can easily provide us with some useful
information about the average energies and widths of the ISGMR of
nuclei extending from the stability to the drip lines.

The shell structure of nuclei near the drip lines considerably differs
from the structure of stable nuclei. However, it should be pointed out
that, in principle, the average properties of the giant resonances are
not appreciably influenced by the pairing correlations at least for
medium and relatively heavy subshell closed nuclei when calculated
with Skyrme forces \cite{R15,R16}. Thus for a fast estimate of some
general trends of the average properties of the ISGMR, we will
restrict ourselves to nonrelativistic Skyrme HF calculations using the
uniform filling approximation, in which particles occupy the lowest
single particle orbits from the bottom of the potential.

Semiclassical methods like the Thomas-Fermi (TF) theory and its
extensions (ETF) which include $\hbar$-corrections \cite{R23} have
proven to be very helpful in dealing with nuclear properties of global
character that vary smoothly with the number of particles $A$ and in
regions where the shell corrections (quantal effects) are small as
compared with the average value provided by the semiclassical
calculations \cite{R20,R21}. Reproduction of the binding energy from
the celebrated Bethe-Weizs\"acker mass formula \cite{R22} is the most
well-known example of this kind. Semiclassical calculations of ETF
type of the average energies of some collective oscillations, in
particular the breathing mode, also reproduce smoothly the quantal RPA
values \cite{R23,R24,R25,R25a1,R25b1} for nuclei close to the
stability line. This can be understood because the ISGMR is a
collective oscillation whose average properties are, in general,
rather insensitive to shell effects which are absent in the
semiclassical approaches of TF type. In this paper we also want to
investigate the ability of the TF approach and its extensions in
reproducing the smooth variation with $A$ of the average energies and
widths of the ISGMR near the drip lines.

The paper is organized as follows: In section II, we review the basic
theory of the sum rule approach applied to obtain the average energy
of the ISGMR using Skyrme forces. The behavior of these average
energies along different chains of isotopes is discussed in section
III paying special attention to the case of Ca isotopes. In section
IV, we study the semiclassical TF and ETF descriptions of the ISGMR
particularly near the drip lines. Finally the summary and conclusions
are given in the last section.

\section{Theory}
The response of the ground state of a nucleus to the action of a multipole 
moment, represented by the operator $Q$, is completely characterized by 
its associated strength function $S(E)$ defined as \cite{R11}
\beq
S(E) = \sum_n {| \langle  n |Q| 0 \rangle  |}^2 \delta (E_n - E),
\label{eq01} \eeq
where $|0\rangle $ and $|n\rangle $ are the normalized ground and excited
states, respectively, and the $E_n$ are the excitation energies.

The moments of the strength function are defined as
\beq
m_k = \int^{\infty}_0 E^k S(E) dE,
\label{eq02} \eeq
where $k$ is an integer. The different moments fulfill the
 inequalities \cite{R11}
\beq
\sqrt{\frac{m_{k+1}}{m_{k-1}}} \ge \frac{m_k}{m_{k-1}} \ge
\sqrt{\frac{m_{k}}{m_{k-2}}},
\label{eq03} \eeq
from where one can define average energies as follows:
\beq
\bar{E_k} = \sqrt{\frac{m_{k}}{m_{k-2}}} \qquad {\rm and} \qquad
\tilde{E_k} = \frac{m_k}{m_{k-1}}.
\label{eq04} \eeq
In addition the square of the variance of the strength is defined as 
\cite{R11}
\beq
\sigma^2 = \frac{m_2}{m_0} - \bigg( \frac{m_1}{m_0} \bigg)^2 \le
\frac{1}{4} \bigg( \frac{m_3}{m_1} - \frac{m_1}{m_{-1}} \bigg).
\label{eq05} \eeq 

With the help of the completeness relation $\sum_n |n\rangle \langle n| = 1$, it can be 
verified that for any positive odd integer $k$ the moments (\ref{eq02}) 
can 
be evaluated as the expectation value in the ground state $|0\rangle $ of some 
commutators which involve the Hamiltonian $H$ and the operator $Q$ 
(assumed to be a hermitian and one-body operator). For instance,
\beq
m_1 = \frac{1}{2} \langle  0 | \big[ Q,\big[ H,Q \big] \big] | 0 \rangle 
\label{eq06} \eeq
and
\beq
m_3 = \frac{1}{2} \langle  0 | \big[ \big[ Q,H \big], \big[ H, \big[ H,Q \big] 
\big] \big]| 0 \rangle ,
\label{eq07} \eeq
which are called the energy weighted and cubic energy weighted sum
rules, respectively \cite{R11}. Equations (\ref{eq06}) and
(\ref{eq07}) as they are written are not useful for practical
calculations because the exact ground state $|0\rangle $ is usually
unknown. However, when the moments are computed within the 1p1h RPA it
can be shown that they coincide exactly with the result obtained by
replacing the actual ground state $|0\rangle $ by the uncorrelated HF
wave function $|{\rm HF}\rangle $ \cite{R11,R11a}.

As usual, to evaluate the energy weighted sum rule $m_1$ we shall
restrict ourselves to an isoscalar single-particle operator $Q= \sum_i
f(r_i)$ and shall neglect the momentum dependent parts of the residual
interaction. In this case one has contributions coming only from the
kinetic energy and at the RPA level one finds:
\beq
m_1 = \frac{1}{2} \langle  {\rm HF} | \big[ Q,\big[ H,Q \big] \big] |{\rm HF} \rangle  =
\frac{\hbar^2}{2m} \langle  {\rm HF} | {(\nabla f)}^2 |{\rm HF} \rangle .
\label{eq08} \eeq                                                                                                    
It has been shown that Eq.\ (\ref{eq08}) also holds for a Skyrme force
due to the $\delta$-character of the $p^2$-terms \cite{R11b}. For the
isoscalar monopole oscillation one has $Q = \sum_{i=1}^A r_i^2$, and
thus the RPA $m_1$ sum rule becomes:
\beq
m_1 = \frac{2 \hbar^2}{m} A \langle  r^2 \rangle , 
\label{eq09} \eeq
where the expectation value of the operator $r^2$ is calculated 
with the HF wave function. 
                                          
\subsection{The scaling approach with Skyrme forces}

As it is known \cite{R11}, the cubic energy weighted moment $m_3$ of
the RPA strength function can be calculated through the scaled
ground-state wave function $\Phi_{\eta}$ which is defined as:
\beq
\vert \Phi_{\eta} \rangle = e^{- i \eta Q_1} \vert \Phi_0 \rangle ,
\label{eq2} \eeq
where $\eta$ is an arbitrary scaling parameter and $Q_1 = i[ H,Q]$,
for a hermitian one-body operator $Q$. Then,
\beq
m_3 = \frac{1}{2} \frac{\partial^2}{\partial \eta^2}
\big[\langle \Phi_{\eta} \vert H \vert \Phi_{\eta} \rangle 
\big]_{\eta=0} .
\label{eq3} \eeq

The $m_3$ moment measures the change of the energy of the
nucleus when the ground-state wave function is deformed according to
(\ref{eq2}). If $Q$ is the monopole
collective operator defined previously, the scaling transformation
(\ref{eq2}) induces a change of scale in the ground-state HF wave function 
$\Phi_0$ conserving its normalization,
i.e. each single-particle wave function varies as:
\beq
\phi_{\eta}^M = e^{3 \tilde{\eta}/2}
\phi_0(e^{\tilde{\eta}} x, e^{\tilde{\eta}} y, e^{\tilde{\eta}} z),
\label{eq4} \eeq
where $\tilde{\eta} = -2 \hbar^2 \eta/m$,
and $\phi_0$ are the single-particle wave functions comprised in
$\Phi_0$.

Under the monopole transformation (\ref{eq4}), the particle,
kinetic and spin densities entering in the Skyrme energy density scale as
\beq
\rho_{\lambda}(\vect{r})= \lambda^3 \rho(\lambda \vect{r}), \qquad
\tau_{\lambda}(\vect{r})= \lambda^5 \tau(\lambda \vect{r}), \qquad
\vect{J}_{\lambda}(\vect{r})= \lambda^5 \vect{J}(\lambda \vect{r}),
\label{eq6} \eeq
where $\lambda=e^{\tilde{\eta}}$.
Inserting these scaled densities in the Skyrme energy density
functional, the scaled energy is obtained as
\beq
E(\lambda)= \lambda^2 T + \lambda^3 E_{\delta}
+ \lambda^5 (E_{fin} + E_{so}) + \lambda^{3 \gamma + 3} E_ {\rho}
+ \lambda E_C ,
\label{eq7} \eeq
where $T$ and $E_C$ are the kinetic and Coulomb energies and $E_{\delta}$,
$E_{fin}$, $E_{so}$, and $E_ {\rho}$ are the different contributions to
the potential energy in the notation of Ref.\ \cite{R11}.
 They are given by
\beq
E_{\delta} = \int d\vect{r} \, \bigg\{ \frac{1}{2} \rho^2 t_0 (1 +
\frac{x_0}{2})
- \frac{1}{2} (\rho_n^2 + \rho_p^2) t_0(\frac{1}{2} + x_0) \bigg\} ,
\label{eq7b} \eeq                 
\beqa
E_{fin} &=& \int d\vect{r} \, \bigg\{ \frac{1}{4} \rho \tau \big[t_1(1
+ \frac{x_1}{2}) + t_2(1
+ \frac{x_2}{2} \big] - \frac{1}{4} (\rho_n \tau_n + \rho_p \tau_p) 
\big[t_1(\frac{1}{2}+x_1) - t_2(\frac{1}{2} + x_2) \big]
\nonumber \\
 &+& \frac{1}{16} (\nabla \rho)^2 \big[3t_1(1 + \frac{x_1}{2}) - t_2(1 + 
\frac{x_2}{2}) \big] \nonumber \\
&-& \frac{1}{16} [(\nabla \rho_n)^2 + (\nabla \rho_p)^2]
 \big[3t_1(\frac{1}{2}+x_1) + t_2(\frac{1}{2} + x_2) \big] \bigg\} ,
\label{eq7a} \eeqa 
\beq
E_{so} = \frac{1}{2} W_0  \int d\vect{r} \, \{ \vect{J} \cdot \nabla
\rho
+ \vect{J}_n \cdot \nabla \rho_n + \vect{J}_p \cdot \nabla \rho_p \},
\label{eq7d} \eeq
\beq            
E_{\rho} = \int d\vect{r} \, \bigg\{ \frac{1}{12} \rho^{ \gamma + 2}
t_3 (1 + \frac{x_3}{2})
- \frac{1}{12} \rho^{\gamma}(\rho_n^2 + \rho_p^2) t_3(\frac{1}{2} + x_3)
\bigg\} .
 \label{eq7c} \eeq

 The stability of the ground-state wave function against a scaling 
transformation implies $E'(\lambda) \vert_{\lambda=1}=0$
(virial theorem) leading to
\beq
2 T + 3 E_{\delta} + 5( E_{fin} + E_{so} ) +(3 \gamma +3) E_{\rho} 
+ E_C = 0.
\label{eq8} \eeq
 According to Eq.\ (\ref{eq3}), the
$m_3$  moment of the isoscalar monopole RPA strength can be written as
 \beq
 m_3 =
 \frac{1}{2} \bigg(\frac{2 \hbar^2}{m} \bigg)^2  \left[ 2T +
6E_{\delta} + 20(E_{fin}
 + E_{so}) + (3 \gamma +3) (3 \gamma + 2) E_{\rho} \right],
\label{eq9} \eeq
 where the scaled ground-state energy (\ref{eq7}) has been employed.
Using Eqs.\ (\ref{eq9}) and (\ref{eq09}), the average
energy of the ISGMR obtained with the scaling approach reads as
\beq
E_M^S = \sqrt{\frac{m_3}{m_1}}. 
\label{eq12a} \eeq

\subsection{Constrained calculation of the giant monopole resonance}  

Let us consider a nucleus, described by a Hamiltonian $H$, under the 
action of a weak one-body field $\eta Q$. Assuming $\eta$ to 
be sufficiently 
small so that perturbation theory holds, the variation of the 
expectation value of $Q$ and $H$ is directly related to the $m_{-1}$
moment \cite{R11},
\beq
m_{-1} = \sum_n \frac{{|\langle n|Q|0\rangle |}^2}{E_n} =
- \frac{1}{2} \bigg[ \frac{\partial \langle Q\rangle }{\partial \eta} \bigg]_{\eta=0} =
 \frac{1}{2} \bigg[ \frac{\partial^2 \langle H\rangle }{\partial \eta^2} 
\bigg]_{\eta=0}.
\label{eq12b} \eeq 

At the RPA level the average energy of the ISGMR can also be
estimated by performing constrained spherical HF calculations, i.e.,
by looking for the HF solutions of the constrained Hamiltonian
\beq
H(\eta) = H - \eta Q ,
\label{eq17} \eeq
where $Q= \sum_{i=1}^A r_i^2$ is the collective monopole operator.
From the HF ground-state solution $\Phi(\eta)$ of the
constrained Hamiltonian (\ref{eq17}), the
RPA $m_{-1}$ moment (polarizability) is computed as
 \beq
 m_{-1}= - \frac{1}{2}
 \bigg[ \frac{\partial}{\partial \eta}
 \langle \Phi(\eta)\vert Q \vert \Phi(\eta) \rangle \bigg]_{\eta=0}
= \frac{1}{2}
\bigg[ \frac{\partial^2}{\partial \eta^2}
\langle \Phi(\eta)\vert H \vert \Phi(\eta) \rangle \bigg]_{\eta=0},
\label{eqsumr2}\eeq
from where another estimate of the average energy of the ISGMR is
given by
\beq
E^C_M = \sqrt{\frac{m_1}{m_{-1}}} .
\label{eq19} \eeq

In the following we will refer to the average energies provided by the
scaling method (\ref{eq12a}) and to the constrained HF calculations
(\ref{eq19}) as the scaled and constrained energies, respectively.
Due to the fact that the RPA moments fulfill the relations
$\sqrt{m_3/m_1} \ge m_1/m_0 \ge \sqrt{m_1/m_{-1}}$ \cite{R11}, the 
values of the average energies
$E^S_M$ and $E^C_M$ are an upper and lower bound of the mean
energy of the resonance, and their difference is related with the
variance of the strength function (resonance width):
\beq
\sigma = \frac{1}{2} \sqrt{ ( E^S_M )^2 - ( E^C_M )^2}.
\label{eq21} \eeq
In the RPA formalism, $\sigma$ is the escape width; that
includes the Landau damping but not the spreading width coming from
 more complicated excitation mechanisms. Therefore,
the width estimated with (\ref{eq21}) can be significantly lower than 
the experimentally measured value \cite{R25}.                                                 

\section{Numerical results}

In our study of the excitation energies
of the giant monopole resonances in the sum rule approach
we consider the isotopes of some proton magic nuclei, namely O, Ca, Ni, Zr 
and Pb, from the proton  up to the neutron drip lines. 
The calculations are
performed using the SkM$^*$ force. 
The results are presented in the following manner. First we 
discuss in detail the behavior of the ISGMR along the isotopic Ca chain. 
There 
are two reasons for that. First of all, the shell effects in the 
$m_{-1}$ sum rule are specially important along this chain. 
Second, the sum rule estimates for 
some isotopes ($^{34}$Ca and $^{60}$Ca) can be contrasted with presently 
existing full RPA calculations \cite{R15,R15a} in order to learn which kind 
of information can be derived from the simpler sum rule approach. 
In the light of 
these discussions in Ca isotopes, we analyze the ISGMR 
in the remaining (O, Ni, Zr and Pb) isotopic chains. 

\subsection{The sum rule approach in calcium isotopes}

In Figure 1 we display the scaled ($E_M^S$) and constrained ($E_M^C$)
estimates of the average excitation energy of the monopole oscillation
as a function of the mass number $A$ from the proton to the neutron
drip lines. In this figure, the filled symbols correspond to closed
subshell nuclei, whereas the open symbols correspond to open shell
nuclei populated according to the uniform filling approximation. The
predicted scaled and constrained energies of these open shell nuclei
are thus for orientation purposes. Along the Ca chain the scaled
monopole energies decrease rather smoothly when the number of
neutrons moves from the stable nuclei towards the neutron drip line.
The constrained energies also show a similar global tendency. However,
two important differences can be observed as compared with the scaled
energies. On the one hand, the reduction of $E^C_M$ near the neutron
drip line is more pronounced than that in the case of scaled energies.
The constrained energies $E^C_M$ also decrease for the proton rich
members of the chain while for these nuclei the scaled energies
$E^S_M$ remain close to the corresponding values in stable nuclei. On
the other hand, relatively strong changes of the constrained energies
can be observed at some subshell closures, in particular for the ones
corresponding to neutron numbers $N=32$ and $N=34$ which are reached
once the orbits $2p_{3/2}$ and $2p_{1/2}$ are completely filled. The
resonance width along the chain, estimated with Eq.\ (\ref{eq21}) is
also displayed in Figure 1. It increases near the neutron and proton
drip lines but also shows noticeable oscillations due to the shell
effects at $N=32$ and $N=34$ in the middle of the Ca chain far from
the drip lines.

The upper panel of Figure 2 displays the $m_3$ sum rule as a function
of the mass number $A$ along the chain of Ca isotopes. From this
figure, it can be seen that $m_3$ smoothly grows with increasing $A$
due to the global character of this sum rule [see Eq.\ (\ref{eq9})].
More interesting is the sum rule $m_3$ divided by $A$, which is shown
in the inset of the same panel. The ratio $m_3/A$ is roughly constant
for the stable nuclei and smoothly decreases when approaching the
neutron and proton drip lines. This behavior can be understood in
terms of the finite nucleus incompressibility $K_A$ \cite{R10} which
is proportional to $m_3/A$. The value of $K_A$ can also be estimated
through its leptodermous expansion \cite{R10}:
\beq
K_A = K_{\infty} + K_{sf} A^{-1/3} +K_{vs} I^2
+K_{coul} Z^2 A^{-4/3} + \ldots ,
\label{eq20} \eeq
where $K_{\infty}$ is the nuclear matter incompressibility and
$K_{sf}$, $K_{vs}$, and $K_{coul}$ are the surface, symmetry, and
Coulomb corrections which in the scaling model are negative
\cite{R10}. Near the neutron drip line, the surface ($K_{sf}
A^{-1/3}$) and Coulomb ($K_{coul} Z^2 A^{-4/3}$) terms decrease in
absolute magnitude and tend to make $K_A$ larger with growing $A$, but
the contribution of the symmetry term ($K_{vs} I^2$) dominates due to
the large values of the neutron excess $I=(N-Z)/A$ and, in total, the
finite nucleus incompressibility $K_A$ is reduced. Near the proton
drip line, all of the three correction terms to $K_A$ are larger in
absolute magnitude than those for stable nuclei, and thus the finite
nucleus incompressibility again decreases. However, this effect near
the proton drip line is smaller than that near the neutron drip line
due to the smaller $\vert I \vert$ values, at least in the case of the
Ca isotopes.

In the middle panel of Figure 2 we display the $m_1$ RPA sum rule for
the chain of Ca isotopes as well as their contribution coming
separately from protons and neutrons. As can be seen, the proton
contribution to $m_1$ is roughly constant due to the fact that the
proton rms radius only changes very little along the whole chain.
Contrarily, the neutron contribution to $m_1$ increases almost
linearly with increasing number of neutrons from the proton drip line
up to $^{70}$Ca from where it starts to grow faster following the
trend of the rms radius of the neutron densities near the neutron drip
line. Due to this behavior of the proton and neutron contributions,
the full $m_1$ as a function of $A$ behaves almost as its neutron
contribution along the chain. It should be pointed out that near the
proton drip line the proton and neutron contributions are rather
similar and both contribute equally to $m_1$, while near the neutron
drip line the $m_1$ sum rule is basically given by its neutron
contribution. 

The $m_{-1}$ sum rule for the chain of Ca isotopes is displayed in the
lower panel of Figure 2 together with their proton and neutron
contributions. From this figure, one can see that the neutron
contribution to $m_{-1}$ increases almost linearly with the mass
number $A$ from the proton drip line till $^{48}$Ca and from $^{54}$Ca
to $^{70}$Ca with sudden changes from $^{48}$Ca to $^{54}$Ca and from
$^{70}$Ca up to the neutron drip line nucleus $^{76}$Ca (shown in the
inset). In contrast, the proton contribution to $m_{-1}$ remains
roughly constant except from $^{40}$Ca to the proton drip line
($^{34}$Ca) where it slightly increases. This behavior of the $m_{-1}$
sum rule together with the one exhibited by $m_1$ sum rule discussed
previously, explains the global trends shown by the constrained
average energy $E^C_M$ of the ISGMR along the Ca chain displayed in
Figure 1 from the proton to the neutron drip lines.

To obtain more insight about the neutron contribution to the $m_1$ and
$m_{-1}$ sum rules we display in Figure 3 the mean square radius of
each neutron orbit $nlj$ entering in the HF wave function of the
ground state of the Ca nuclei along the isotopic chain. They can be
contrasted with the neutron mean square radii of the same nuclei,
shown by the thick horizontal lines. To help the discussion and for
further purposes, the neutron and proton single-particle energy levels
of the Ca isotopes are displayed in Figures 4 and 5 respectively. In
these figures we have also included some quasi-bound levels owing to
the centrifugal (neutrons) or centrifugal plus Coulomb (protons)
barriers which simulate possible positive energy single-particle
resonant states. From Figure 3 it can be seen that the deeper orbits
almost give the same mean square radius along the whole isotopic chain
in agreement with the fact that the more bound energy levels do not
change very much from the proton to the neutron drip line (see Figure
4). Beyond $^{48}$Ca, neutrons start to get accommodated in levels
whose binding energies are relatively small and consequently the rms
radii of the corresponding orbits increase. It is important to note
that, in particular, the $2p_{3/2}$ and $2p_{1/2}$ orbits have a very
large mean square radius in spite of the fact that their binding
energies are not very small specially for the heaviest isotopes. The
mean square radius of these $p$ orbits is larger than the one
corresponding to the less bound level $1f_{5/2}$ and the one of the
$2p_{1/2}$ orbit is similar to the one of the $1g_{9/2}$ orbit which
lies higher in the energy spectrum. Near the neutron drip line, the
very lightly bound level $2d_{5/2}$ is occupied and its wave function
extends rather further increasing substantially the mean square radius
of $^{76}$Ca (the contribution of the $2d_{5/2}$ level to the mean
square radius is shown by a triangle in the Figure 3 in the reduced
scale as indicated). This fact produces a noticeable departure of the
linear behavior exhibited by the $m_1$ sum rule near the neutron drip
line. 

In Figure 6 we display the contributions $m_{-1}[nlj]$ of a single
nucleon in the occupied neutron orbits to the total $m_{-1}$ sum rule
for each Ca isotope. From this figure one can see that the more bound
neutron levels give almost the same single-particle contribution to
$m_{-1}$ in all the nuclei of the Ca chain. The behavior of the $m_
{-1}$ sum rule from $^{48}$Ca onwards to the neutron drip line is
governed by the outer orbits $2p_{3/2}$, $2p_{1/2}$, $1f_{5/2}$,
$1g_{9/2}$ and $2d_{5/2}$ which give the largest contribution to
$m_{-1}$ from $^{52}$Ca to $^{76}$Ca. The single-particle
contributions to $m_{-1}$ from the orbits $2p_{3/2}$ and $1g_{9/2}$
are still roughly constant, however, strong changes can be observed in
$m_{-1}[nlj]$ for the $2p_{1/2}$ and $1f_{5/2}$ orbits. A very large
$m_{-1}[nlj]$ value comes from the outermost very lightly bound
orbital $2d_{5/2}$ in the drip line nucleus $^{76}$Ca (shown as a plus
sign in the Figure in the reduced scale as indicated). The large
single-particle contribution of the $2p_{1/2}$ and $2d_{5/2}$ orbits
to $m_ {-1}$ is due to the extension of their corresponding wave
functions  beyond the core of the nucleus, so that neutrons in such
orbitals  are much softer to pick up than neutrons accommodated in
inner orbitals producing the strong enhancement of the total $m_ {-1}$
sum rule.

Let us now discuss the enhancement of the $m_{-1}$ sum rule near the
proton drip line. As it is known \cite{R17a}, the depth of the proton
single particle-potential for protons decreases when the number of
neutrons in a isotopic chain decreases. Thus the outermost bound
proton levels for nuclei near the proton drip line (see Figure 5) have
little binding energy, their corresponding HF wave functions extend
far from the core of the nuclei and consequently protons in these
orbits are softer against pick-up increasing the contribution to the
total $m_ {-1}$ sum rule. This is just the case of the $1d_{3/2}$
orbit for protons in Ca isotopes (see Figure 5), which passes from
$-7.5$ MeV in $^{40}$Ca to $-1.1$ MeV in the proton drip line nucleus
$^{34}$Ca and gives the largest contribution to $m_{-1}$ in this case.

\subsection{Connection with the RPA strength function}

As mentioned, the sum rules $m_1$, $m_3$, and $m_{-1}$ computed from
the HF ground state by means of Eqs.\ (\ref{eq09}), (\ref{eq9}), and
(\ref{eqsumr2}), are the {\em exact} 1p-1h RPA value \cite{R11}.
Actually, the self-consistent HF sum rules provide a practical means
to check the accuracy of an RPA calculation of the strength function
$S(E)$, provided that both the sum rule and the RPA calculation are
performed in the same conditions. However, because of their
complexity, it often happens that the RPA calculations are not fully
self-consistent. For instance, the Coulomb and/or spin-orbit residual
interactions may be absent in the RPA response. It has been recently
shown \cite{AS} that when the RPA calculations are performed
self-consistently, the sum rules extracted from the RPA strength
function are quite close to the values obtained in the HF sum rule
approach.       

RPA calculations for $^{34}$Ca and $^{60}$Ca were carried out some
time ago \cite{R15}. The extracted ISGMR mean energies were
$E^S_M=22.5$ MeV and $E^C_M=17.8$ MeV for $^{34}$Ca, and $E^S_M=18.8$
MeV and $E^C_M=15.4$ MeV for $^{60}$Ca. Although there are some
differences, they compare well with the values from our sum rule
calculation: $E^S_M=22.0$ MeV and $E^C_M=17.3$ MeV for $^{34}$Ca, and
$E^S_M=18.3$ MeV and $E^C_M=15.0$ MeV for $^{60}$Ca. Our values for
the resonance widths are 6.8 MeV in $^{34}$Ca and 5.3 MeV in
$^{60}$Ca, which coincide with the RPA values obtained in Ref.\
\cite{R15}. The systematic deviation of about 0.5 MeV of the RPA
average energies reported in Ref.\ \cite{R15} from our sum rule
results may possibly be attributed to a deficiency in full
self-consistency.

The information contained in the energy spectra of the nuclei as well
as in the values of the sum rules allows one to infer important
properties of the strength function $S(E)$. Looking at Figure 10 of
Ref.\ \cite{R15}, one can see that for $^{34}$Ca there is a low energy
bump in the region around 5--13 MeV\@. It is mainly due to transitions
from the less bound proton levels ($1d_{3/2}$ and $2s_{1/2}$) to the
continuum. These transitions enhance the $m_{-1}$ sum rule that weighs
the low energy part of $S(E)$. The peak seen at high energies, in the
19--25 MeV region, receives contributions from the $1p \rightarrow 2p$
(resonant) transition of protons as well as from the $1p \rightarrow
2p$ and $1s \rightarrow 2s$ transitions of neutrons. Concerning
$^{60}$Ca (Figure 12 of Ref.\ \cite{R15}), the low energy bump around
4--12 MeV is again mainly related to transitions to the continuum from
the less bound levels, which in this case are due to neutrons. For
this nucleus the last neutron in the $1f_{5/2}$ orbit is bound by 3.4
MeV which is in agreement with the threshold energy. The high energy
peaks around 17 MeV and 21 MeV can be related to the $2s \rightarrow
3s$ (virtual) and to the $1d_{5/2} \rightarrow 2d_{5/2}$ (resonant)
transitions for protons and to the $1p \rightarrow 2p$, $1d_{5/2}
\rightarrow 2d_{5/2}$ and $1d_{3/2} \rightarrow 2d_{3/2}$ (resonant)
transitions for neutrons. It should be pointed out that these
transitions between single-particle levels, strictly speaking, should
be related to peaks of the unperturbed strength, but they also appear
when the RPA correlations are included. They are, however, slightly
shifted to lower energies in general. With this experience in hand, we
discuss from a qualitative point of view the most important trends of
the RPA monopole response which can be expected for some
representative nuclei of the Ca isotopic chain.

For $^{52}$Ca and $^{54}$Ca, a low energy bump in $S(E)$ starts to be
developed due to transitions from $2p_{3/2}$ and $2p_{1/2}$ levels to
the continuum. Although the strength in the low energy region of $S(E)$
should be small due to the relatively large binding energy of the last
filled levels of neutrons of these nuclei (6.7 MeV and 4.1 MeV for
$^{52}$Ca and $^{54}$Ca respectively), the large rms radii of these
orbits favor a large overlap between the wave functions of these bound
levels and the levels in the continuum \cite{R17b} which increases the
strength at these energies. This fact is supported, on average, by the
enhancement of the $m_{-1}$ sum rule in passing from $^{48}$Ca to
$^{54}$Ca.  This confirms our assumption about the key role of the
$2p_{3/2}$ and $2p_{1/2}$ orbits on the sudden increase of $m_{-1}$.
In passing from $^{60}$Ca to $^{70}$Ca, the increase of $m_1$ and
$m_{-1}$ is basically a size effect which explains the almost similar
values of $E^C_M$ in both nuclei. Here one should also expect a low
energy bump shifted towards lower energies due to transitions to the
continuum  from the $1g_{9/2}$ orbit which is bound by 2.5 MeV. For
$^{76}$Ca,  $m_{-1}$ strongly increases due to the effect of the
$2d_{5/2}$ orbit (bound by only 0.84 MeV) which has a very large rms
radius. Although $m_1$ slightly increases, the large change of
$m_{-1}$ explains the dramatic reduction of $E^C_M$ at the neutron
drip line for the Ca isotopes. For this nucleus one could also expect
a large low energy bump dominated by the transition to the continuum
from the $2d_{5/2}$ level. The structure in $S(E)$ should also contain
broad high energy peaks due to transitions from the deeply bound
states to the unoccupied bound or resonant levels, coming from both
neutrons and protons. From this it appears that the knowledge of the
spectrum of a nucleus as well as the $m_1$ and $m_{-1}$ sum rules can
provide useful information that allows one to sketch the behavior of
the strength distribution $S(E)$.

\subsection{Transition densities}

In this subsection we will discuss the two following transition 
densities \cite{R13}:
\beq
{\delta \rho}_{+1} (\vect{r}) = - \frac{1}{2 A \langle  r^2\rangle } \big[ 3 \rho (\vect{r})
+ r \rho' (\vect{r}) \big]
\label{eq98} \eeq
which is the Tassie transition density corresponding to the scaling 
transformation and 
\beq
{\delta \rho}_{-1}(\vect{r}) = - \frac{1}{2 m_{-1}}
\left. \frac{\partial \rho_{\eta}(\vect{r})}
{\partial \eta } \right\vert_{\eta=0}
\label{eq99} \eeq
which is the transition density in the constrained case. 
In Eq.\ (\ref{eq98})
$\rho(\vect{r})$ is the ground-state particle density and $\langle r^2\rangle $ the mean 
square radius of the nucleus. 
In Eq.\ (\ref{eq99}) $\rho_{\eta}(\vect{r})$ is the 
particle density built up with the HF single-particle wave functions 
that are solutions 
of the constrained Hamiltonian (\ref{eq17}) and $m_{-1}$ is the inverse 
energy weighted sum rule calculated using (\ref{eqsumr2}). With the chosen 
normalization in Eqs.\ (\ref{eq98}) and (\ref{eq99}), it is ensured that 
when a single state exhausts the sum rule, then ${\delta 
\rho}_{+1}(\vect{r})={\delta \rho}_{-1}(\vect{r})$. 

Figure 7 displays the neutron (upper panel) and proton (middle panel) 
contributions to the constrained transition densities (\ref{eq99}) 
multiplied by $r^2$ for some Ca isotopes between the proton and neutron 
drip 
lines, namely $^{34}$Ca, $^{48}$Ca, $^{54}$Ca, $^{70}$Ca and $^{76}$Ca. The 
neutron contribution to $r^2 {\delta \rho}_{-1}$ strongly depends on the 
neutron number of the considered isotope. The outer bump of the neutron 
contribution broadens and shifts to larger values of $r$ when
the nuclei approach
the neutron drip line; this is  basically due to the neutrons in the outermost 
orbits which are loosely bound and extend very far from the center 
of the nucleus. For example, in Figure 7 one can see the 
effect of the $2p_{3/2}$ and $2p_{1/2}$ orbits in $^{54}$Ca where the outer 
bump is clearly shifted with respect to that of the $^{48}$Ca nucleus. The 
fact that the rms radius of the $1f_{5/2}$ and $1g_{9/2}$ orbits are 
similar to the ones of the $2p_{3/2}$ and $2p_{1/2}$ orbits (see Figure 3), 
is also reflected in the transition density of the nucleus $^{70}$Ca whose
external bump is located, roughly, at the same position as that of the one 
corresponding to $^{54}$Ca. The very lightly bound $2d_{5/2}$ orbit has a 
very large rms radius (see Figure 3) and consequently the external bump of 
the neutron contribution to the constrained transition density is pushed to 
very large distances. The proton contribution to the constrained transition 
density displayed in the middle panel of Figure 7 shows a completely different 
pattern. On the one hand, both the inner and outer bumps are located 
approximately at the same positions for all the considered Ca 
isotopes, which is in agreement with the fact that the proton
 rms radius along a 
given isotopic chain remains approximately constant. On the other hand, the 
proton contribution to the total transition density decreases with  
increasing mass number . This is consistent with the fact that proton 
contributes proportionally less than neutrons to the $m_{-1}$ sum rule 
in neutron drip line nuclei as can be seen from the lower panel of Figure 
2.

The lower panel of Figure 7 compares the Tassie and the constrained transition 
densities for the drip line 
nuclei $^{34}$Ca and $^{70}$Ca. For both nuclei the bumps of $r^2 {\delta 
\rho}_{+1}$ and $r^2 {\delta \rho}_{-1}$ are located at the same place. 
However, the shapes of both transition densities differ relatively between 
them.
The strong fragmentation of the RPA strength in Ca isotopes, in 
particular in the drip line nuclei, can be gauged  by the large 
enhancement of the resonance width in these nuclei (see Figure 1). 

\subsection{Other isotopic chains}

With the experience gained from this analysis of Ca isotopes, let us 
discuss the more salient features found in other isotopic chains. 
Figure 8 displays the scaled ($E^S_M$) and constrained ($E^C_M$) energies of 
the ISGMR as well as the resonance width for the O, Ni, Zr and Pb 
isotopic 
chains. The filled and open symbols correspond to subshell closed and open 
nuclei respectively. 
The scaled energies along these isotopic chains qualitatively behave in a 
similar way as those in Ca isotopes. In general, $E^S_M$ is roughly
constant along the considered chains from the proton drip line to the 
stable nuclei and then shows a moderate downwards trend while 
approaching the neutron drip line. On the contrary, the constrained 
energies exhibit again a sizeable reduction near the neutron drip line and 
a moderate decrease near the proton drip line. One exception to this 
general behavior is $^{12}$O which shows a noticeable decrease 
of the scaled and constrained energies at the proton drip line. The 
reasons are similar to those discussed in the case of the $^{34}$Ca,
but more accentuated because of the smaller nuclear charge.
Another exception is the heavy Pb chain where the constrained energy
$E^C_M$ smoothly diminishes from the proton to the neutron drip lines 
showing a similar behavior to that of the scaled energy for these isotopes.   
As a consequence, for light and medium isotopic chains (O, Ni and Zr)
the
ISGMR width shows an enhancement near the proton and neutron drip lines 
(somewhat more pronounced at the neutron drip line), whereas the resonance 
widths
for the Pb isotopes remain approximately constant along the whole
chain. Thus, the resonance width in the different
isotopic chains shows a downwards tendency with increasing atomic number, 
which suggests a smaller fragmentation and a more collective character
in the ISGMR of heavy nuclei such as the Pb isotopes.

We now discuss the $m_1$ and $m_{-1}$ sum rules in O, Ni, Zr and Pb 
isotopes. In Figures 9 and 10
we display $m_1/A^{5/3}$ and $m_{-1}/A^{7/3}$, respectively, for the
aforementioned isotopic chains. The scaling factors $A^{-5/3}$ and
$A^{-7/3}$ remove the size dependence of 
the total $m_1$ and $m_{-1}$ sum rules \cite{R25}.
 From Figure 9 it can be seen that the neutron contribution to 
$m_1/A^{5/3}$ increases and the proton contribution decreases with growing 
mass number $A$. However, the total $m_1/A^{5/3}$ shows a rather constant 
behavior as a function of $A$
for all the considered isotopic chains, except maybe for some drip line
nuclei.  This means that the enhancement 
of the $m_1$ sum rule with $A$ displayed in Figure 2 for Ca isotopes up to 
$^{70}$Ca is basically due to a size effect except near the drip line nuclei. 
 The $A$-scaled $m_{-1}/A^{7/3}$ sum rule displayed in Figure 10 for 
the O, Ni, Zr and Pb chains shows a rather constant behavior in the region 
of stable nuclei. It clearly increases near the neutron drip lines and also 
for the proton drip line of oxygen.
It shows the softness against pick-up 
 of the nucleons from the outermost lightly  
bound levels thus giving the large 
enhancement of the $m_{-1}$ sum rule in the case of drip line nuclei.

From the sum rule approach we can also obtain some information about the 
RPA strength distributions for O, Ni, Zr, and Pb isotopes. In Ref.
\cite{R25a}
the isoscalar monopole  RPA strength for the neutron drip line nucleus 
$^{28}$O was studied. We can analyze this nucleus on the basis of the
sum rule approach. To help the discussion, we report in Table 1 
the neutron and proton single-particle energy levels (including the 
quasi-bound ones) as well as the mean square radius of each bound orbit 
for this nucleus obtained with the SkM$^{*}$ force. In 
this case the last occupied neutron level ($1d_{3/2}$) is bound by only 
1.8 MeV. As discussed in Ref.\ \cite{R25a}, the RPA strength develops a
low energy bump due to transitions to the continuum from the $1d_{5/2}$, 
$2s_{1/2}$, and $1d_{3/2}$ levels, and a high energy peak due to the
neutron
transition $1p_{3/2} \rightarrow 2p_{3/2}$ (resonant) and the proton 
transitions from the $1s_{1/2}$, $1p_{1/2}$, and $1p_{3/2}$ levels to
the unoccupied (but bound) $2s_{1/2}$, $2p_{1/2}$, and $2p_{3/2}$
levels, respectively. This structure is actually encoded in the
sum rule approach used here. Looking at Figure 10, a noticeable 
enhancement of the scaled $m_{-1}/A^{7/3}$ sum rule for $^{28}$O can be 
seen in  qualitative agreement with the large amount of strength in the 
low energy part of $S(E)$. 
To be more quantitative, we also report in Table 1 the contribution to
$m_{-1}$ coming from a single neutron in each occupied orbit. 
The largest part of the $m_{-1}$ sum rule comes from contributions of the 
$1d_{3/2}$, $2s_{1/2}$ and $1d_{5/2}$ neutron levels, which are 
responsible for 74\%, 11\% and 8\% of the total $m_{-1}$ sum rule 
(263.022 MeV$^{-1}$ fm$^4$), respectively. On the other hand, using the 
$\langle r^2\rangle $ values reported in the same Table, it is found that the 
contribution of these neutron levels to the total $m_1$ sum rule (2.586 
$\times$ 10$^4$ MeV fm$^4$) is 23\%, 11\% and 25\%, pointing out again
their relevance in the low energy part of $S(E)$.

The single-nucleon contributions to $\langle r^2\rangle $ and to the
$m_{-1}$ sum rule reported in Table 1 can also provide another
information about $S(E)$. The energy corresponding to the maximum of
the unperturbed strength due to transitions of the outermost nucleons
to the continuum can be estimated for each orbit as the square root of
the ratio of their single nucleon contributions to $m_1[nlj]$ and
$m_{-1}[nlj]$. From the values reported in Table 1, this estimate
gives 5.6 MeV and 9.6 MeV for the $1d_{3/2}$ and $2s_{1/2}$ neutron
levels, in qualitative agreement with the position of the maxima in
the unperturbed strength displayed in Figure 1 of Ref.\ \cite{R25a}.

For the Ni isotopes the situation is qualitatively similar to that
found for the Ca isotopic chain, although there are some differences.
First, the enhancement of the polarizability
shown by Ca isotopes when the $2p_{3/2}$ and $2p_{1/2}$ orbits are
occupied
($^{52}$Ca and $^{54}$Ca respectively) is appreciably attenuated in
the corresponding Ni isotopes ($^{60}$Ni and $^{62}$Ni). 
 This is due to the fact that for Ni isotopes the 
$1f_{7/2}$ orbit for protons is filled and thus neutrons occupying the
$2p_{3/2}$ and $2p_{1/2}$ levels are more bound and have smaller rms 
radii, as compared with the corresponding Ca isotopes. They thus
contribute less to the $m_{-1}$ sum rule as can be realized from
Figure 10 where no significant enhancement of the scaled
$m_{-1}/A^{7/3}$ sum rule is observed.
Second, in Ca isotopes the strong enhancement of the $m_{-1}$ sum rule 
and a moderate rise in $m_1$ start to be appreciable once the 
$1g_{9/2}$ orbit has been occupied (which corresponds to $^{70}$Ca) and the 
very lightly bound $2d_{5/2}$ level starts to be populated (see Figure 2).
Something similar happens for Ni isotopes beyond $^{78}$Ni where the
$2d_{5/2}$, $3s_{1/2}$ and $2d_{3/2}$ orbits become slightly bound and 
contribute largely to the enhancement of the $m_{-1}$ (strong) and
$m_1$ (moderate) sum rules producing the decrease of the constrained 
energy and the enhancement of the width of the ISGMR in nuclei between 
$^{78}$Ni and the drip line nucleus $^{90}$Ni. However, the single nucleon 
contribution to the $m_1$ and $m_{-1}$ sum rules critically depends 
on its binding energy and increases very fast when the binding 
energy approaches zero. In our uniform filling approach and using the 
SkM$^*$ force, the last occupied level in the drip nucleus $^{76}$Ca 
($2d_{5/2}$) is bound by only 0.85 MeV while in the Ni drip nucleus 
$^{90}$Ni the 
outermost levels $2d_{5/2}$, $3s_{1/2}$ and $2d_{3/2}$ are bound by 3.45, 2.26 
and 1.09 MeV respectively. These differences in the binding energies of the 
last occupied levels explain why the decrease in the constrained
energy and the enhancement of the width of the ISGMR is more important
in the Ca isotopes than in
the Ni isotopes near the neutron drip line.
    
We next discuss the chain of the Zr isotopes. For the $^{122}$Zr
nucleus where the neutron shell $N=82$ is closed, the binding energy
of the last bound orbit ($1h_{9/2}$) is 5.01 MeV with the SkM$^*$
force. Some neutron orbits belonging to the $N=126$ closed shell,
namely the $2f_{7/2}$, $3p_{3/2}$ and $3p_{1/2}$ orbitals, become very
lightly bound in the $^{130}$Zr, $^{134}$Zr and $^{136}$Zr isotopes
and thus increase largely the $m_1$ and $m_{-1}$ sum rules in these
nuclei close to the neutron drip line. In the uniform filling
approximation, such a scenario can be appreciated in Figures 9 and 10
from the departure from the roughly constant value of the $A$-scaled
$m_1/A^{5/3}$ and $m_{-1}/A^{7/3}$ sum rules in subshell closed Zr
isotopes beyond $^{122}$Zr. The strong enhancement of the $m_{-1}$ sum
rule combined with the moderate rise in the $m_1$ sum rule produces
the dramatic decrease of the constrained energy $E^C_M$ and the large
increase of the resonance width near the neutron drip line which can
be observed in Figure 8.

The situation is different in the heavy Pb isotopic chain. In this
case the neutron drip line is reached once the $N=184$ shell is
completely occupied (the last filled orbit $3d_{3/2}$ is bound by 2.58
MeV). In the lighter Ca, Ni and Zr isotopic chains, the shell gap
above the $N=50$ and $N=82$ closed shells is strongly reduced when
approaching the neutron drip line (and even disappears in the case of
the Ca isotopes \cite{R1e}) as compared to those of the stable nuclei
of these chains. Due to this fact, in nuclei near the neutron drip
line some levels of the next major shell can still be bound before
reaching it. However, in neutron-rich Pb isotopes the shell closure at
$N=184$ is more robust and there are no bound levels belonging to the
next major shell in the neutron drip nucleus. The $A$-scaled
$m_1/A^{5/3}$ sum rule is practically constant along the whole chain
and $m_{-1}/A^{7/3}$ shows a moderate increasing tendency near the
neutron drip line (see Figures 9 and 10). The combination of these two
effects causes the constrained energy and the energy width of the
ISGMR in Pb neutron rich isotopes vary only slightly as compared to
the corresponding values in the stable nuclei of the chain which can
be seen in Figure 8.

We have treated all nuclei in the spherical approximation. This is
justified to a large extent because of the semi-magic character of the
isotopic chains considered. However, some of the isotopes may happen to be
deformed, specially in the Zr chain. Since more than two decades ago,
there is experimental evidence \cite{BU,GA} that the isoscalar monopole
strength in deformed nuclei shows an additional lower energy peak when
compared to that of spherical nuclei. This is interpreted as an effect of
the coupling between the ISGMR and the isoscalar giant quadrupole
resonance in deformed nuclei that broadens and splits the ISGMR strength
into two components \cite{YO,IT}. We believe that this effect may be
enhanced in deformed isotopes away from the valley of stability, because
the broader distribution of the resonance strengths near the drip lines
possibly would contribute to an increase of the coupling between the two
excitation modes.

 \section{Semiclassical Extended Thomas-Fermi calculations}

As mentioned in the introduction, it is known that for stable nuclei
semiclassical approximations of Thomas-Fermi (TF) type and its
extensions (ETF)
used together with the sum rule approach provide a good estimate of
the ISGMR excitation energies at least for stable nuclei
\cite{R23,R24,R25,R25a1,R25b1}. In this section we want to study
whether these semiclassical approximations are also able to follow the
average trend of the quantal results in regions away from the
stability line.

To describe semiclassically (at TF or ETF levels) the ISGMR excitation
energies with the scaling method, we shall calculate first the
ground-state neutron and proton densities. This is done by replacing
the quantal kinetic energy density $\tau = \sum_{i=1}^A {\vert
\vect{\nabla} \phi_i \vert}^2$ by the semiclassical one at TF level
$\tau_{TF} = \sum_{q=n,p} 3/5{(3 \pi^2)}^{2/3} \rho_q^{5/3}$ or at ETF
level including $\hbar^2$ or $\hbar^4$ corrections \cite{R23,R24} into
the Skyrme energy density functional and then solving the variational
Euler-Lagrange equations for the neutron and proton densities $\rho_n$
and $\rho_p$, respectively \cite{R24}. The ETF method introduces
$\hbar^2$ ($\hbar^4$) corrections on top of the simple TF kinetic
energy through an expansion of second (fourth) order gradients of the
particle density which include non-local contributions coming from the
spin-orbit potential and the effective mass. These $\hbar$ corrections
modify the asymptotic behavior of the self-consistent solution of the
energy density and give a semiclassical density profile which averages
the shell oscillations of the quantal HF density in the bulk and
reproduces its fall-off at the surface. When $\hbar^4$ corrections are
taken into account, the semiclassical energy of the ground state
becomes roughly similar to the one obtained using the more cumbersome
Strutinsky average method \cite{R24}.

In the semiclassical approaches, the neutron (proton) drip line is
reached when the neutron (proton) chemical potential vanishes. We have
checked \cite{R29} that the neutron drip lines for Ca and Pb at TF
level using the SkM$^*$ force correspond to neutron numbers of 48 and
195 which are in reasonable agreement with the HF predictions (56 and
184 respectively). This fact points out that nuclei near the drip
lines can be studied through semiclassical methods of the ETF type. If
the TF or ETF densities are used to compute (\ref{eq8}), the virial
theorem is, of course, strictly fulfilled due to the self-consistency. The
semiclassical sum rules $m_3$ and $m_1$, which are needed to obtain
the excitation energy of the ISGMR in the scaling method, are calculated
using again Eqs.\ (\ref{eq9}) and (\ref{eq09}) but with the
semiclassical self-consistent neutron and proton densities instead of the
HF ones.

As it happens in the quantal case, the excitation energy of the ISGMR can 
also be estimated semiclassically 
by performing TF or ETF constrained calculations. In this case one 
has to minimize the constrained  semiclassical energy 
\beq
\int d\vect{r} [ {\cal H}_{TF,ETF} - \eta r^2 \rho_{\eta} ] =
E(\eta) - \eta \int d\vect{r} r^2 \rho_{\eta} ,
\label{eq21a} \eeq                        
where the 
quantal kinetic energy has been replaced by the TF or ETF equivalents.
 In order to obtain the semiclassical $m_{-1}$ sum rule, use is 
made of Eq.\ (\ref{eq12b}) but replacing the HF expectation values by the 
semiclassical ones. 

Thus the semiclassical average excitation energies of the ISGMR are
obtained from the RPA sum rules but with the HF expectation values
entering in them replaced by the TF or ETF ones. The semiclassical
excitation energies of the ISGMR calculated with the Skyrme SkM$^{*}$
force using the scaling method and performing semiclassical
constrained calculations are displayed in the upper panel of Figure 11
in comparison with the exact RPA average energies $\bar{E}_3=
\sqrt{m_3/m_1}$  and $\bar{E}_1= \sqrt{m_1/m_{-1}}$ (which are the
quantal scaled ($E^S_M$) and constrained ($E^S_M$) energies described
in Section 2) along the Ca isotopic chain from the proton to the
neutron drip lines. The semiclassical predictions of the excitation
energies displayed in Figure 11 are calculated at pure TF level
(dashed-dotted lines) and at ETF level including $\hbar^2$ (solid lines) and
$\hbar^4$ (dashed lines) corrections. From this Figure it can be seen that the
general trends of the excitation energies $\bar{E}_3$ of the ISGMR are
well  reproduced by all the considered semiclassical approximations
along the whole isotopic chain. The agreement between the RPA
$\bar{E}_3$ energies and their semiclassical counterpart improves by
increasing the order of the $\hbar$-corrections as can be seen in the
upper panel of Figure 11. If $\hbar^4$-corrections are included in the
ETF calculation, then the RPA $\bar{E}_3$ energies are almost
perfectly reproduced.

As far as the scaling approach is concerned, a first conclusion of
this analysis is that the semiclassical approach to the RPA
$\bar{E}_3$ (scaling) energies of the ISGMR agrees very well with the
corresponding quantal RPA values not only for stable nuclei but also
for nuclei near the drip lines, specially if $\hbar^4$ corrections are
included in the semiclassical calculation. This fact points out that
the role of shell effects is almost negligible in the estimate of the
average energy $\bar{E}_3$ of the ISGMR obtained from the $m_3$ and
$m_1$ sum rules on the one hand, and it justifies the use of the
liquid-drop like expansion of the finite nucleus incompressibility
(\ref{eq20}) along the full isotopic chain on the other hand.

However, the situation is different when the ISGMR excitation energy
is estimated by performing constrained calculations. In this case the
constrained TF calculations give excitation energies which lie very
close to the values obtained with the scaling method for Ca isotopes
and thus fail in reproducing the RPA $\bar{E}_1$ energies in nuclei
far from stability. Consequently, in the simple TF approximation, the
predicted resonance width is very small and cannot reproduce the
behavior of the RPA estimates of the width (\ref{eq21}) along the
chain as it can be seen in the lower panel of Figure 11, where the RPA
widths and their semiclassical equivalents are displayed. The
agreement between the RPA average energies $\bar{E}_1$ and their
semiclassical estimates considerably improves when $\hbar$ corrections are
considered explicitly in the calculation of the semiclassical densities.
The ETF-$\hbar^2$ average energy of the ISGMR obtained by
semiclassical constrained calculations qualitatively describes the
behavior of the RPA results. A more quantitative agreement, at least
from the proton drip to stable nuclei, is found when the average
excitation energies are obtained through ETF-$\hbar^4$ calculations.
Of course, the effects induced by some specific orbitals in the RPA
$\bar{E}_1$ energies in Ca isotopes cannot be recovered by the
semiclassical approaches. However, they nicely average the RPA
results, the average being better when the $\hbar^4$-corrections are taken
into account in the calculation. A similar situation is found in the
estimates of the width of the ISGMR resonance along the isotopic
chain, where the semiclassical approaches average again the RPA values
as it can be seen in the lower panel of Figure 11.

A comparison between the RPA average excitation energies $\bar{E}_3$
and $\bar{E}_1$ and their semiclassical estimates is displayed in
Figure 12 for the Pb isotopic chain. As it happens for Ca isotopes,
the semiclassical scaled energies vary smoothly with the mass number
$A$ and reproduce quite well the RPA $\bar{E}_3$ average energies from
the proton to the neutron drip lines, at least for the subshell closed
isotopes displayed in the figure. For Pb isotopes, the RPA $\bar{E}_1$
average energies show a rather smooth decreasing tendency with increasing 
mass number $A$ which is enhanced near the neutron drip line. As it
happens for Ca isotopes, the semiclassical TF constrained energies lie
close to the scaled ones and are unable to describe the strong
decreasing of $\bar{E}_1$ near the neutron drip line. However, when
$\hbar$-corrections are added, the semiclassical ETF estimates nicely
reproduce the RPA $\bar{E}_1$ average energies, specially if
$\hbar^4$-corrections are taken into account. As has been discussed before,
the ISGMR of Pb isotopes exhibit a much more collective behavior and
single-particle  effects are much less important than in Ca isotopes
even near the drip lines. Thus it is not completely surprising that
the semiclassical approximations of ETF type are able to reproduce
accurately not only the $\bar{E}_3$ average energies but also the
$\bar{E}_1$ ones.

\section{Summary and conclusions}

In this paper we have analyzed the variation of average properties
(mean energies and resonance widths) of the isoscalar giant monopole
resonance along the isotopic chains between the proton and neutron
drip line for nuclei with magic atomic number from O
to Pb. These average energies are obtained within the RPA sum rule
approach using the SkM$^*$ interaction. For each nucleus we have
calculated the RPA cubic ($m_3$) and inverse energy ($m_{-1}$)
weighted sum rules. The $m_3$ sum rule is obtained by means of a
scaling transformation of the self-consistent HF neutron and proton
densities, while $m_{-1}$ is computed by performing
constrained Hartree-Fock calculations.
For a Skyrme force the energy-weighted sum rule $m_1$ which is
required for evaluating the average energies, is simply proportional
to the mean square radius of the Hartree-Fock particle density.

The scaled energies $\bar{E}_3$ along the isotopic chains show general
trends which are rather independent of the atomic number. The scaled
estimate of the average excitation energy of the ISGMR shows a
downwards tendency as one moves towards the neutron drip line from the
stable nuclei. This fall-off of the scaled energy is smooth due to the
moderate growing behavior of the $m_3$ sum rule with increasing mass
number which is compensated by the stronger enhancement of the $m_1$
sum rule. The latter weighs more the tail of the density distributions
and consequently is more important in nuclei near the neutron drip
line. This reduction of the scaled estimate of the ISGMR average
excitation energy at the neutron drip line is more noticeable in small
and medium mass nuclei and its relative importance decreases with
increasing atomic number.

A similar decreasing tendency of $\bar{E}_3$ near the proton drip line
appears only for very light nuclei such as the O isotopes, and it is
not observed in heavier systems. The reason lies in the fact that for
proton-drip line nuclei, the Coulomb barrier prevents their
single-particle wave functions from extending much from the center of the
nucleus, thus prohibiting an enhancement of the rms radius of the
density which in turn would reduce the scaled average excitation
energies of these nuclei. Due to the global character of the scaled
estimate of the ISGMR average energies that depend on the nuclear
densities, single-particle effects play a minor role on these energies
even for drip line nuclei. In this regard, the macroscopic description
of the finite nucleus incompressibility based on a leptodermous
expansion is still valid and allows for a qualitative understanding of the
behavior of the scaled estimates of the ISGMR average energies even
near the drip lines.

The global behavior of the average energies of the ISGMR estimated
through constrained calculations is, in general, similar to the one
exhibited by the scaled energies. These average energies are in
general smaller in drip line nuclei than those in stable ones. The
effect is much more pronounced for the constrained energies calculated
near the neutron drip line, in contrast to those calculated in the
scaling approximation. Near the neutron drip line the constrained
estimate of excitation energy of the ISGMR clearly deviates from the
empirical $A^{-1/3}$ law known for stable nuclei. The reason for this
behavior of the constrained estimate near the drip lines is due to the
fact that the single-particle effects are much more important in this
case than in the scaling calculation. Nuclei near the drip lines are
characterized by neutrons (protons) occupying very lightly bound
levels whose corresponding wave functions extend very far from the
core of the nucleus, specially in
the case of neutron rich nuclei owing to the absence of Coulomb barrier. 
Neutrons occupying these orbits have larger single-particle neutron mean 
square radius and at the same time are much softer against pick-up 
than those filling more bound orbits. 
Consequently, both the energy and the inverse energy weighted sum rules 
are enhanced, this effect being more important in the latter than in the 
former. This explains the sizeable decrease of the constrained
estimate of the
average energy of the ISGMR in nuclei near the neutron drip lines. This 
effect is particularly important in the Ca and Zr isotopic chains where the 
corresponding neutron drip nuclei ($^{76}$Ca and $^{136}$Zr respectively) 
have the last occupied orbit bound by less than 1 MeV. It should also be 
pointed out that for the particular nuclei $^{52}$Ca and $^{54}$Ca which are 
not neutron drip nuclei, we find a noticeable enhancement of the $m_1$ 
and $m_{-1}$ sum rules due to the fact that for these specific nuclei 
the neutron $2p_{3/2}$ and $2p_{1/2}$ single-particle wave 
functions extend very 
far from the core although their binding energies are not particularly small.

The RPA sum rule description of some of the average properties of the
ISGMR has some clear limitations as compared with a full RPA
calculation of the response function. Thus only some global trends of
the RPA strength can be inferred from our calculation. For instance,
the reduction of the scaled and constrained energies near the drip
lines should be associated with the appearance of a noticeable RPA
strength in the low energy region as suggested by the enhancement of
the $m_{-1}$ sum rule. As in a full RPA calculation, nuclei near the
drip lines are found to have a large resonance width in our approach,
but the mean energy of the resonance cannot be determined very
precisely because the scaled and constrained energies, which are upper
and lower bounds of the mean energy, are largely separated. This large
width near the drip lines can be due to a very broad resonance as well
as to a fragmentation of the strength distribution.

We have also investigated the ability of semiclassical approximations
of the TF type and its extensions including $\hbar$-corrections (ETF)
for describing the ISGMR near the drip lines within the sum rule
approach. The semiclassical sum rules are obtained by using the full
quantal RPA expressions but with the HF expectation values replaced by
the semiclassical ones at the TF or the ETF levels. The semiclassical
estimates of the average scaled and constrained energies are free from
any shell effects. Thus some of the discussed trends, as for instance
the enhancement of the $m_1$ and $m_{-1}$ sum rules when a particular
orbital is occupied, are only taken into account on the average in the
semiclassical description. This means that the quantal RPA scaled and
constrained energies may be spread around the values provided by the
semiclassical calculations. In nuclei near the drip lines the pure TF
approximation fails in reproducing, even on the average, the global
trends of the RPA constrained energies and thus the widths of the
ISGMR, due to the poor description of the nuclear surface. However,
when the $\hbar^2$-order corrections, and those of $\hbar^4$ order,
are included in the semiclassical calculation, the ETF description of
the surface of the nuclei agrees better with the HF one and the RPA scaled and
constrained energies are nicely averaged by the corresponding semiclassical
counterpart. 

It is known that a right description of nuclei near the drip lines has to 
take into account pairing correlations, in particular for open shell nuclei. 
The inclusion of pairing in RPA, i.e. the quasiparticle RPA, has a 
noticeable effect on the strength of the response of the drip line nuclei 
to  external fields \cite{QRPA}; however its impact on the average 
excitation energies is much less. Thus, in the present approximation
we did not consider pairing and assumed the uniform
filling approach in order to obtain 
an insight into the general trends of the ISGMR in exotic nuclei near
the drip lines.
In a next step, the application of the sum rule
approach in the quasiparticle RPA should be probed in order to properly 
take into account the influence of pairing 
correlations on the average properties of the 
ISGMR. Investigations in this direction are in progress.

\acknowledgments{Useful discussions with J. Martorell, G. Col\`o, N.
Van Giai, and S. Shlomo are gratefully acknowledged. This work has
been partially supported by grants BFM2002-01868 (DGI, Spain, and
FEDER) and 2001SGR-00064 (DGR, Catalonia).}

\pagebreak


%
\pagebreak

%
%
\begin{table}
\begin{center}
\caption{Neutron and proton single-particle energy levels (in MeV)
with the occupancy (given within the round brackets) for
$^{28}$O. The mean square radius $\langle r^2\rangle$ (in fm$^2$) and
the single-nucleon contribution
to the $m_{-1}$ sum rule of each orbital (in MeV$^{-1}$fm$^4$)
are also given.}
\vspace{0.5cm}

\begin{tabular}{|c|c|c|c|c|c|c|c|}
\hline \hline
 Neutron State & Energy (n) & $\langle r^2\rangle$ & $m_{-1}[nlj]$
    & Proton State  & Energy (p)& $\langle r^2\rangle$ & $m_{-1}[nlj]$
    \\ \hline
$1s_{1/2}$ & $-31.920$ (1) & 5.085 & 0.3093
            & $1s_{1/2}$ &$-44.132$ (1) & 5.433 & 0.3505 \\
$1p_{3/2}$ & $-19.920$ (1) & 8.701 & 0.9664
            & $1p_{3/2}$ &$-31.726$ (1) & 8.786 & 1.0010 \\
$1p_{1/2}$ & $-14.936$ (1) & 9.194 & 1.5875
            & $1p_{1/2}$ &$-27.060$ (1) & 9.068 & 1.3510 \\
$1d_{5/2}$ &  $-8.641$ (1) & 12.962 & 3.6757
            & $1d_{5/2}$ &$-19.389$ (0) &---&--- \\
$2s_{1/2}$ &  $-5.997$ (1) & 16.689 & 15.1465
            & $2s_{1/2}$ &$-15.143$ (0) &---&--- \\
$1d_{3/2}$ &  $-1.780$ (1) & 18.278 & 48.9019
            & $1d_{3/2}$ &$-12.415$ (0) &---&--- \\
$2p_{3/2}$ &   $0.526$ (0) &---&---
            & $1f_{7/2}$ & $-6.665$ (0) &---&--- \\
$1f_{7/2}$ &   $1.731$ (0) &---&---
            & $2p_{3/2}$ & $-2.380$ (0) &---&--- \\
$1g_{9/2}$ &  $10.366$ (0) &---&---
            & $2p_{1/2}$ & $-0.701$ (0) &---&--- \\ \hline
\end{tabular}
\end{center}
\end{table}
%
\newpage
\begin{figure}
\includegraphics[width=0.85\linewidth, angle=0, clip=true]{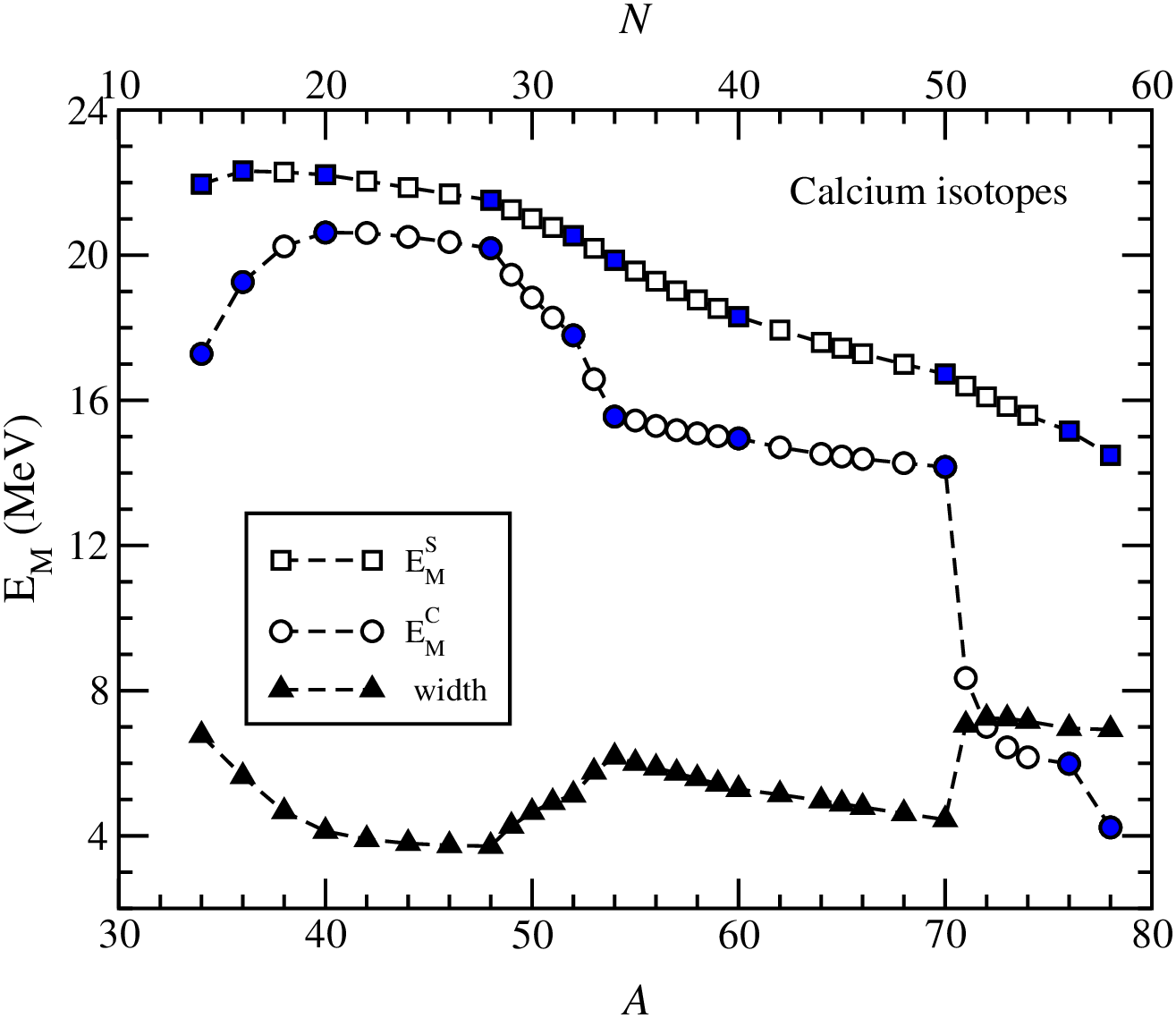}
\caption{\label{Fig1} (Color online)
Excitation energies of the isoscalar giant monopole
         resonance in the isotopic chain of calcium from scaling (squares)
         and constrained (circles) calculations as function of the mass 
         number. The estimate of the resonance width (filled triangles)
         from Eq.\ (\ref{eq21}) is also displayed.
} \end{figure}
\newpage
\begin{figure}
\includegraphics[width=0.85\linewidth, angle=0, clip=true]{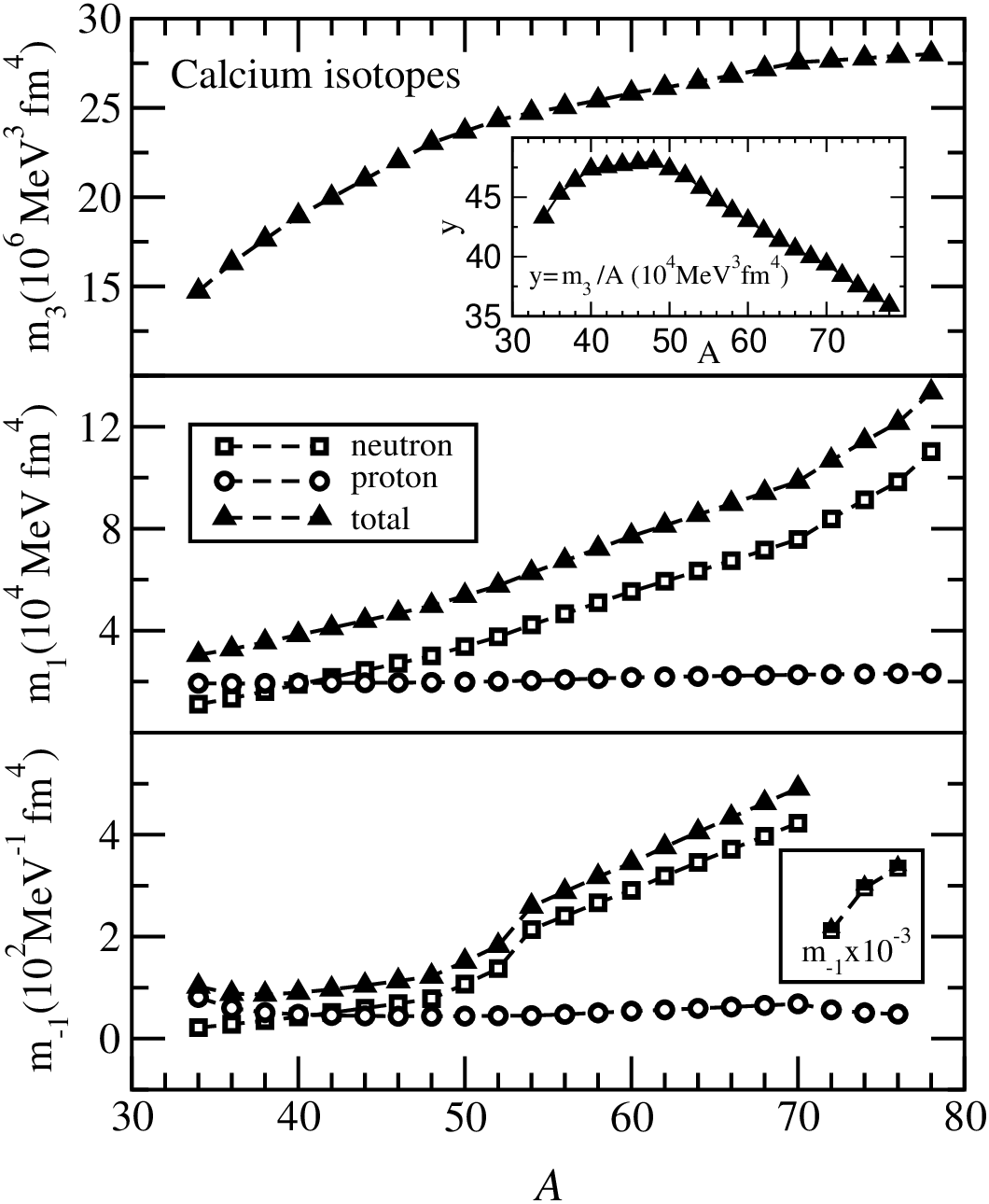}
\caption{\label{Fig2} The $m_3$ sum rule for calcium isotopes as a
        function of the mass number $A$ is shown in the upper panel.
The total $m_1$ and $m_{-1}$ sum rules (filled triangles) alongwith
        the contributions coming from neutrons (squares) and
        protons (circles) are displayed in the middle and lower
panels.} \end{figure}  
\newpage
\begin{figure}
\includegraphics[width=0.85\linewidth, 
angle=0, clip=true]{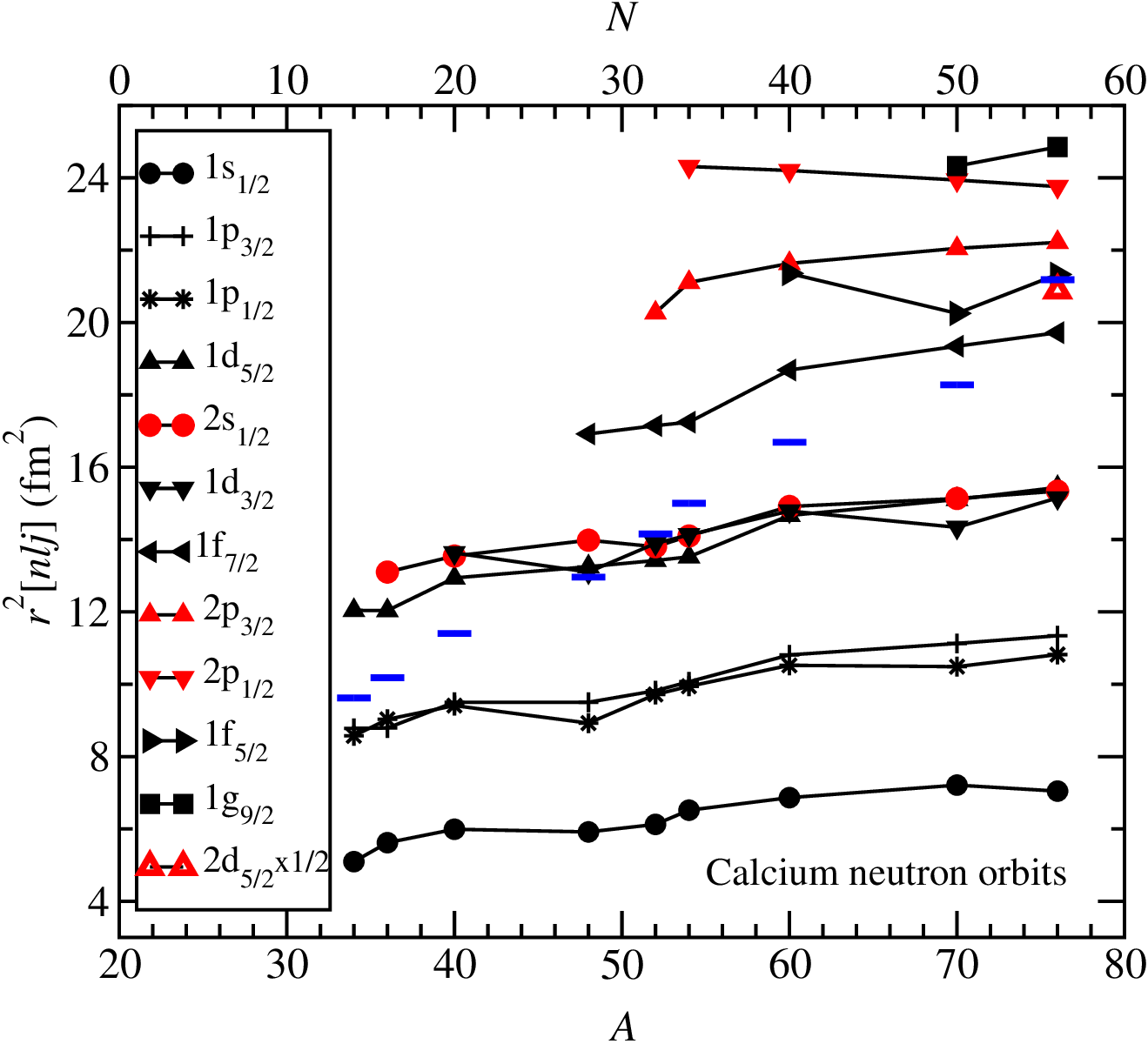}
\caption{\label{Fig3} (Color online) Mean square radii of the
neutron single-particle orbits for the isotopic chain of calcium as a
function of the mass number. The horizontal bars depict the value of
the total neutron mean square radius of each isotope.
}
\end{figure}  
\newpage
\begin{figure}
\includegraphics[width=0.85\linewidth, angle=0, clip=true]{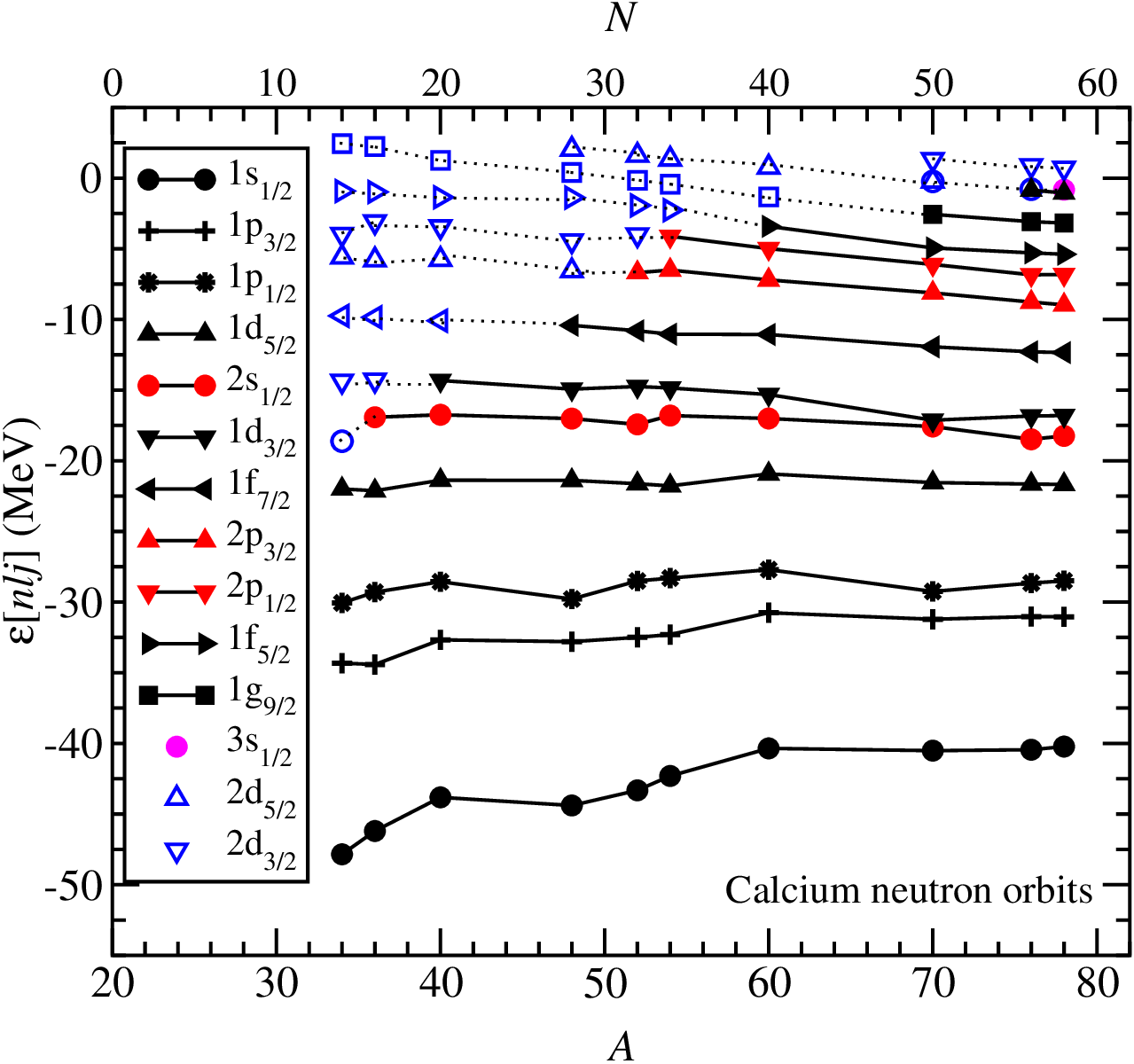}
\caption{\label{Fig4} (Color online) Neutron single-particle energies
for calcium
         isotopes as a function of $A$. Symbols joined by the dotted
         lines correspond to empty levels.}
\end{figure}  
\newpage
\begin{figure}
\includegraphics[width=0.85\linewidth, angle=0, clip=true]{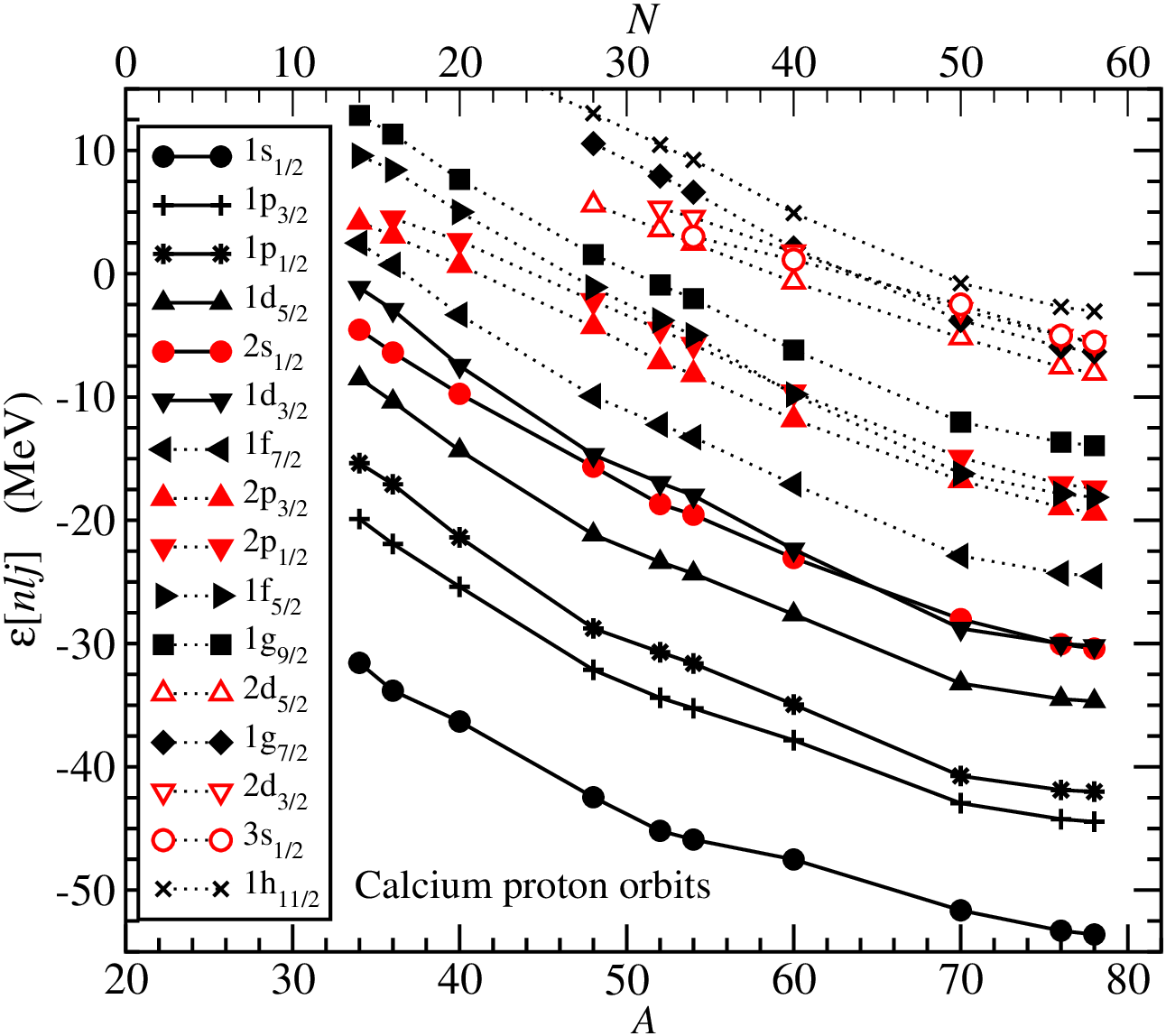}
\caption{\label{Fig5}  (Color online) Same as Figure~\ref{Fig4} but
for protons.} \end{figure}  
\newpage
\begin{figure}
\includegraphics[width=0.85\linewidth, angle=0, clip=true]{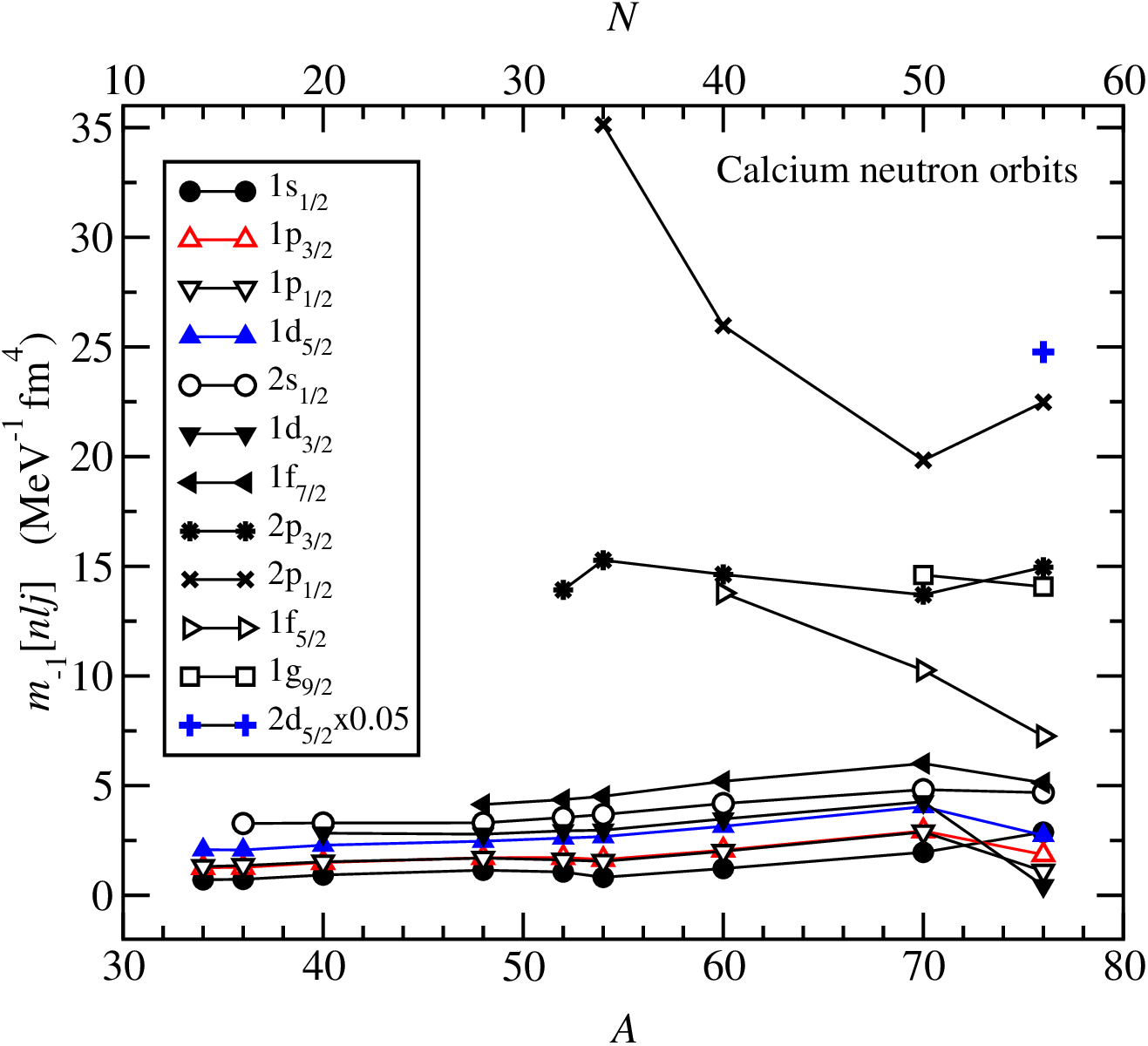}
\caption{\label{Fig6} (Color online) Contributions of a single neutron
in each occupied orbit to the
$m_{-1}$ sum rule for the calcium isotopes.}
\end{figure}  
\newpage
\begin{figure}
\includegraphics[width=0.85\linewidth, angle=0, clip=true]{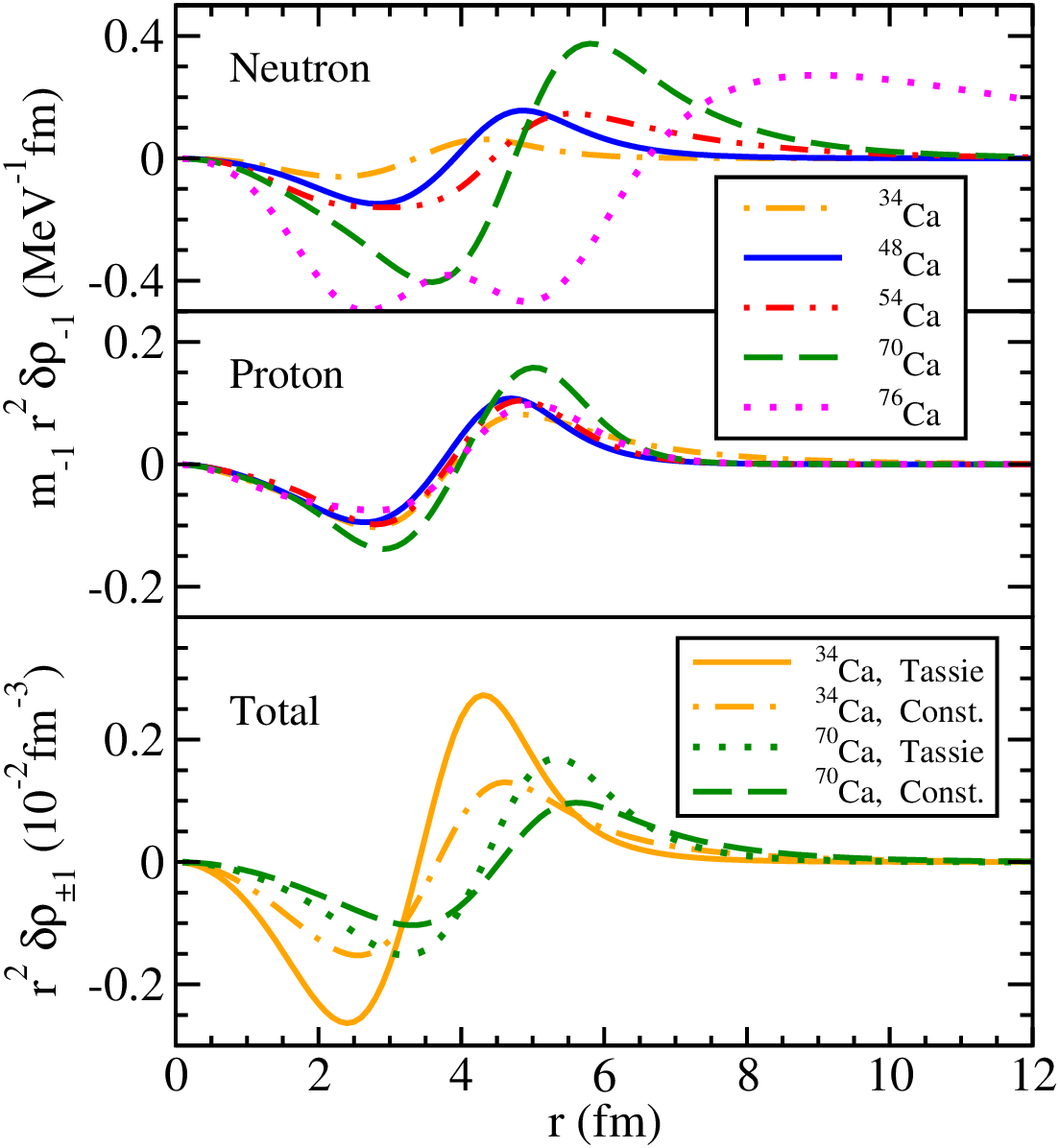}
\caption{\label{Fig7} (Color online) Spatial variation of the
constrained transition densities for neutrons (upper panel) and
protons (middle panel) of some selected Ca isotopes. 
The Tassie and constrained transition densities are compared in
the lower panel. Notice the different vertical scales.}
\end{figure}  
\newpage
\begin{figure}
\includegraphics[width=0.9\linewidth, angle=0, clip=true]{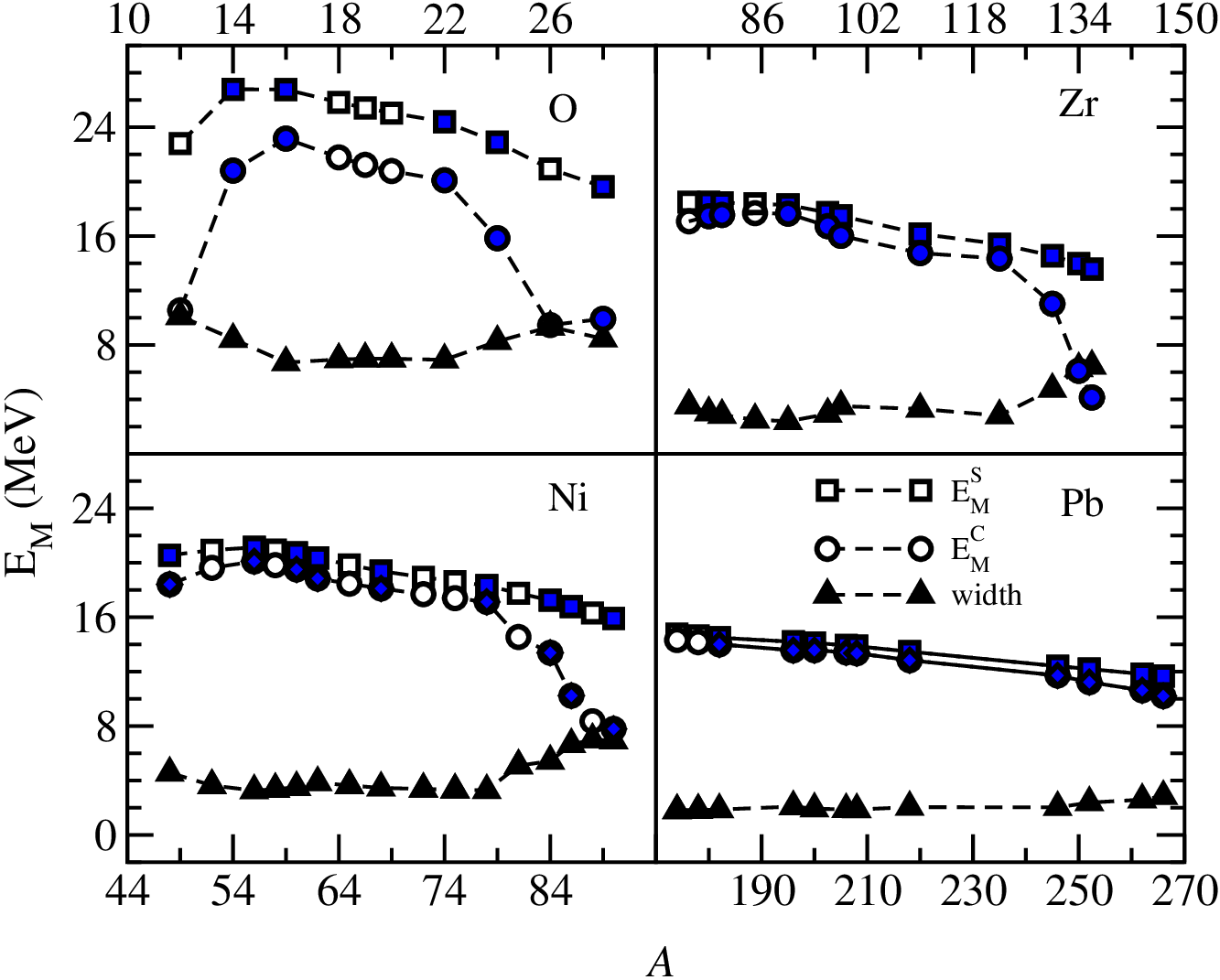}
\caption{\label{Fig8} (Color online) Excitation energy of the ISGMR
from scaling and constrained calculations alongwith the predicted
resonance width for the isotopic chains of O, Ni, Zr, and Pb.}
\end{figure}  
\newpage
\begin{figure}
\includegraphics[width=0.9\linewidth, angle=0, clip=true]{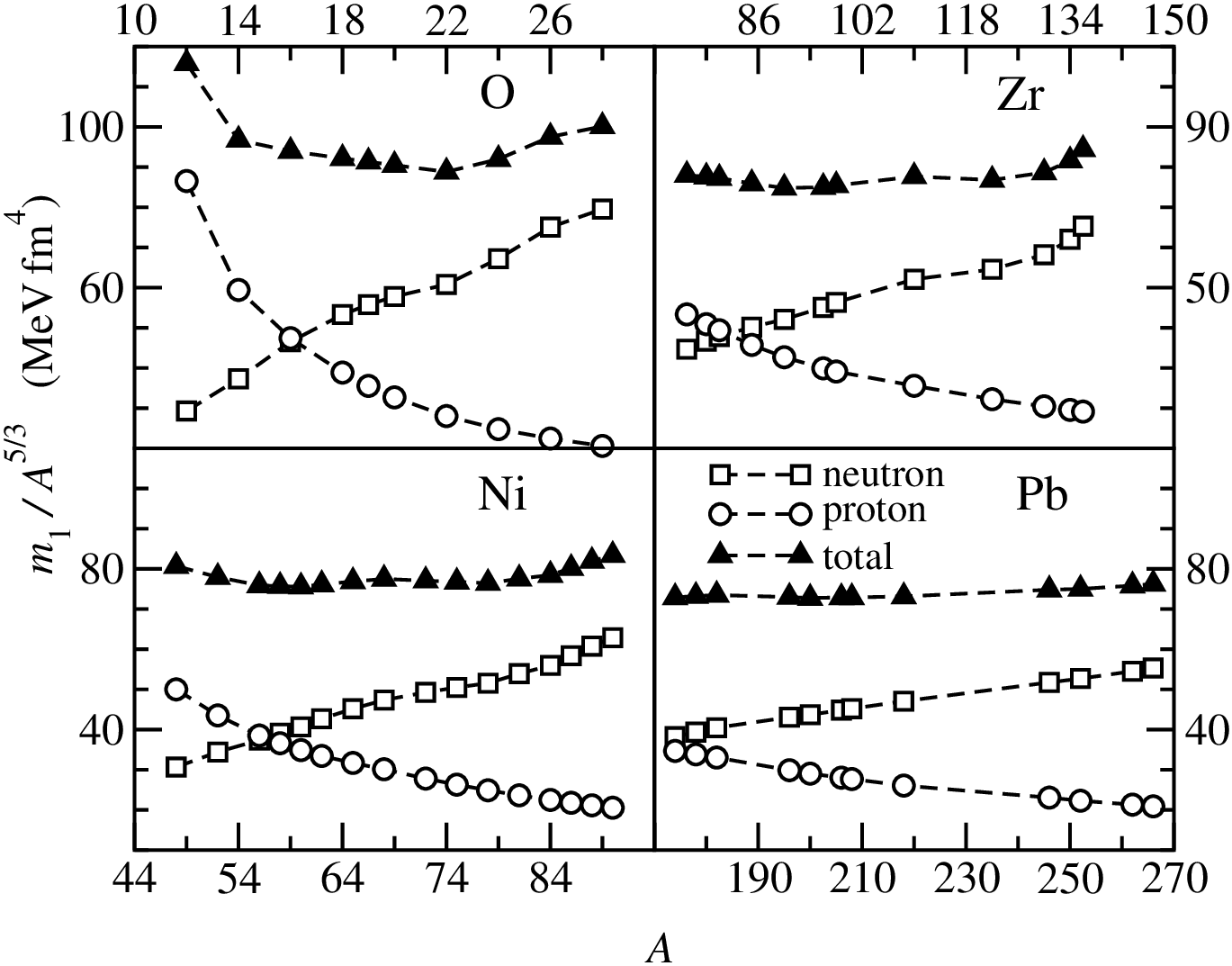}
\caption{\label{Fig9} The $m_1$ sum rule scaled by $A^{-5/3}$ is
displayed
        for the isotopic chains of O, Ni, Zr, and Pb. The neutron
        (squares) and proton (circles) contributions to the total $m_1$
        are also plotted.}
\end{figure}  
\newpage
\begin{figure}
\includegraphics[width=0.9\linewidth, angle=0, clip=true]{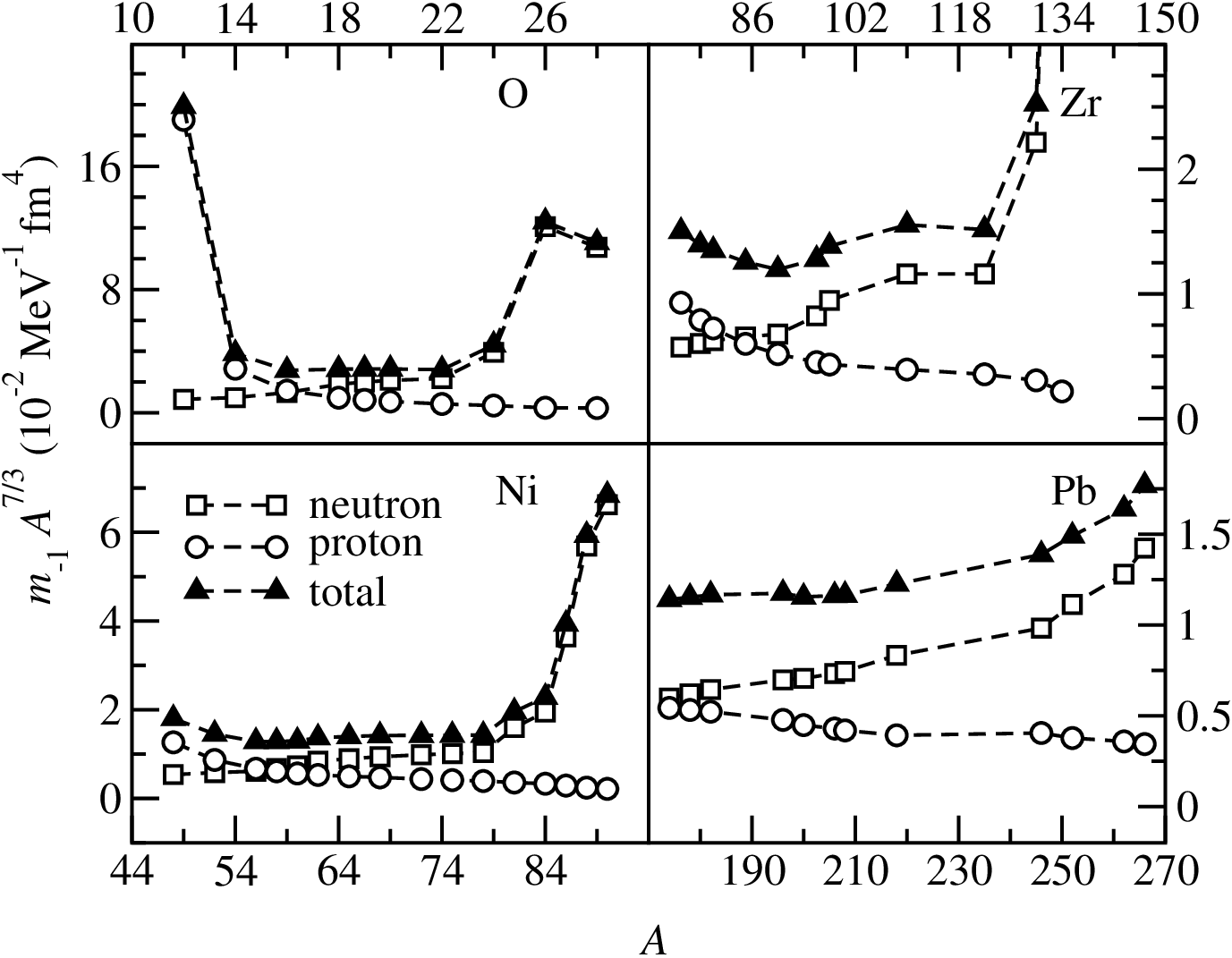}
\caption{\label{Fig10} The $m_{-1}$ sum rule scaled by $A^{-7/3}$ is
displayed
        for the isotopic chains of O, Ni, Zr, and Pb. The neutron
        (squares) and proton (circles) contributions to the total $m_{-1}$
        are also plotted.} 
\end{figure}  
\newpage
\begin{figure}
\includegraphics[width=0.85\linewidth, angle=0, clip=true]{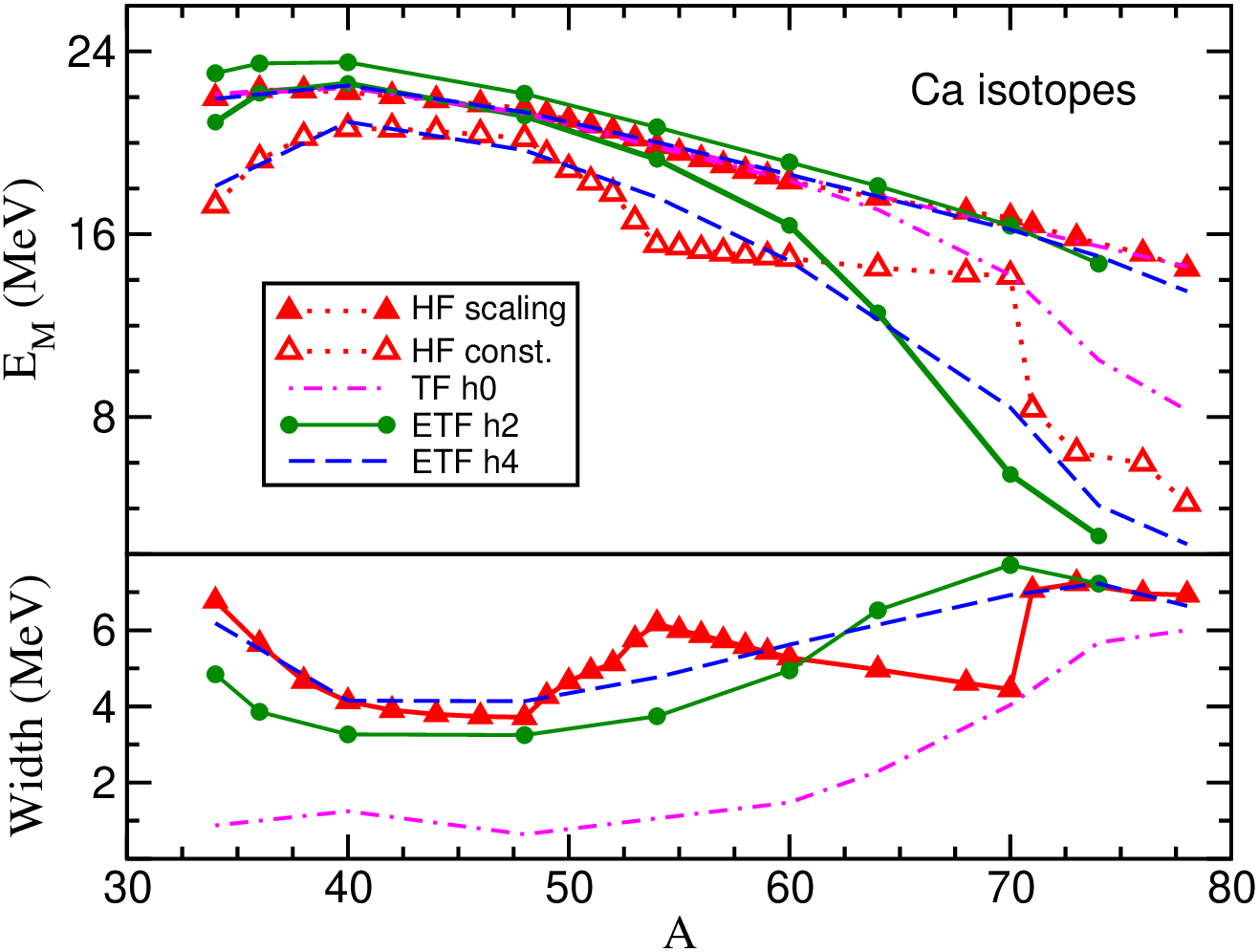}
\caption{\label{Fig11} (Color online) The variation of the excitation
energies of the isoscalar giant monopole resonance with the mass
number $A$ for the Ca isotopic chain is shown in the upper panel.
Semiclassical results of Thomas-Fermi (TF) type including $\hbar$
corrections at different orders, obtained with the scaling and
constrained approaches, are contrasted with the RPA average energies.
The resonance widths are presented in the lower panel.}
\end{figure}  
\newpage
\begin{figure}
\includegraphics[width=0.85\linewidth, angle=0, clip=true]{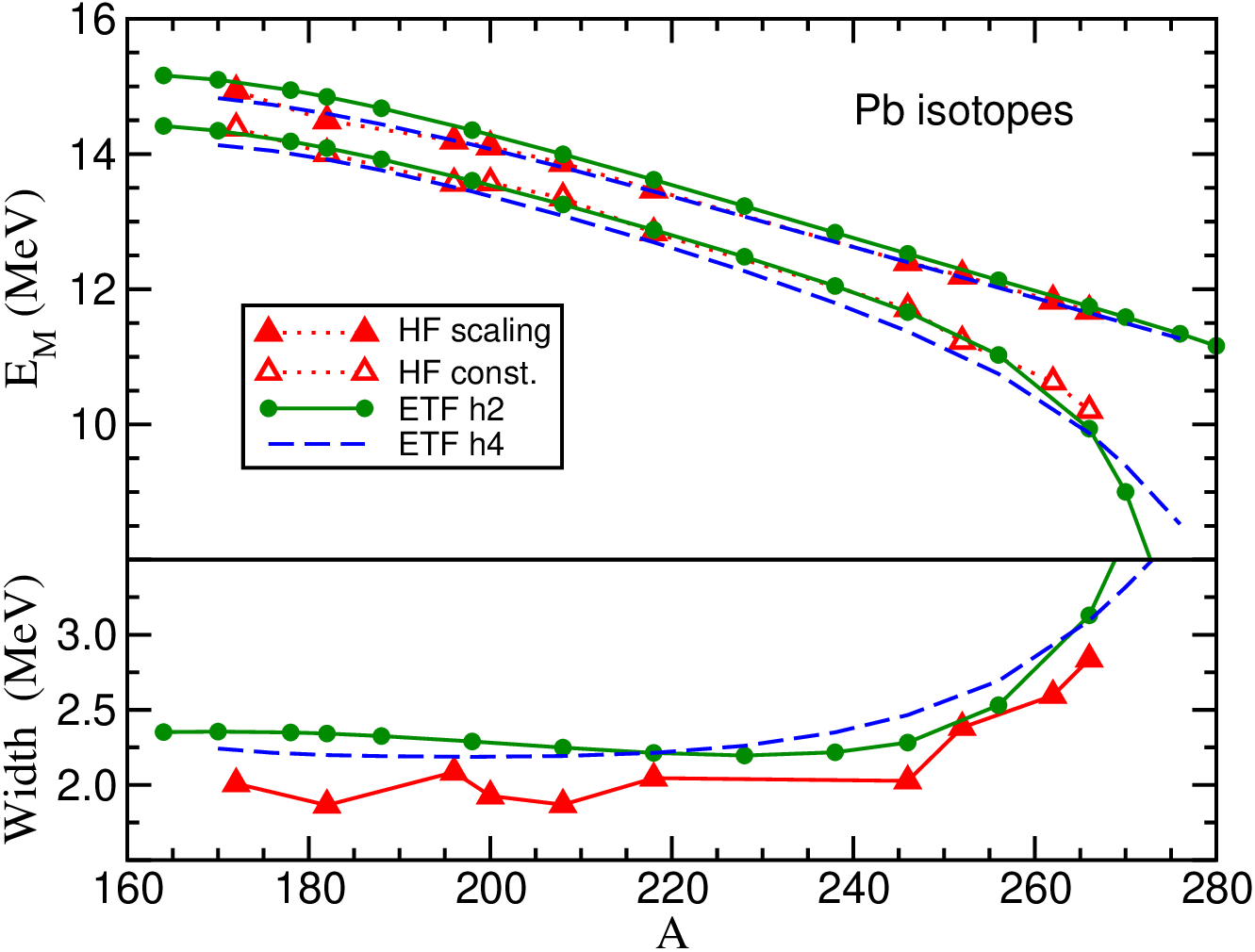}
\caption{\label{Fig12} (Color online) Same as Figure~\ref{Fig11}
but for Pb isotopes.} \end{figure}  
\end{document}